\documentclass{aa}  
\usepackage{titlesec}
\usepackage{multirow}
\usepackage{float}
\usepackage{xcolor}
\usepackage{graphicx}
\usepackage{txfonts}
\usepackage{lipsum}
\usepackage{subcaption}      
\usepackage{lscape}              
\usepackage{placeins}    
\usepackage{hyperref}
\usepackage{pdfcomment}
\usepackage{ulem}
\hypersetup{
  colorlinks=true, 
  breaklinks=true,
  linkcolor=blue, 
  citecolor=blue, 
  filecolor=blue, 
  urlcolor=blue,
  unicode=false, 
  pdftoolbar=true, 
  pdfmenubar=true, 
  pdffitwindow=false, 
  pdfstartview={Fit}, 
  pdftitle={}, 
  pdfauthor={}, 
  pdfsubject={},
  pdfcreator={}, 
  pdfkeywords={},
  pdfnewwindow=true, 
  pdfdisplaydoctitle=true 
}
\defcitealias{robitaille2017}{R17}
\defcitealias{richardson24}{R24}

\makeatletter
\renewcommand\thelinenumber{\relax} 
\let\linenumbers\relax
\let\nolinenumbers\relax
\let\do@mlinenumbers\relax
\let\do@linenumber\relax
\let\@LN@LLerror\@gobble
\makeatother

\begin{document}

   \title{SED Modeling of Young Stellar Objects in the Orion Star Formation Complex}


   \author{Ilknur Gezer\inst{\ref{konkoly},\ref{mtaexcellence}}
          \and
          Gábor Marton\inst{\ref{konkoly},\ref{mtaexcellence}}
          \and          
          Julia Roquette\inst{\ref{inst-ch-obs}}
          \and
          Marc~Audard\inst{\ref{inst-ch-obs}}
          \and
          David Hernandez\inst{\ref{univie}}
          \and
          Máté Madarász\inst{\ref{konkoly},\ref{mtaexcellence}}
          \and
          Odysseas Dionatos\inst{\ref{nhmvie}}
          }

   \institute{
              Konkoly Observatory, Research Centre for Astronomy and Earth Sciences, Hungarian Research Network (HUN-REN), H-1121 Budapest, Konkoly Thege Miklós út 15-17, Hungary\label{konkoly}\\
              \email{marton.gabor@csfk.org}
         \and
              CSFK, MTA Centre of Excellence, Budapest, Konkoly Thege Miklós út 15-17, H-1121, Hungary\label{mtaexcellence}
         \and
              Department of Astronomy, University of Geneva, Chemin Pegasi 51, 1290 Versoix, Switzerland\label{inst-ch-obs}
         \and
              Department of Astrophysics, University of Vienna, Türkenschanzstrasse 17, 1180 Vienna, Austria\label{univie}
        \and
              Natural History Museum Vienna, Burgring 7, 1010 Vienna, Austria\label{nhmvie}
             }
   \date{Received September 30, 20XX}

 
    \abstract
    {One of the key tools for understanding the evolution of young stellar objects (YSOs) is to analyze their spectral energy distributions (SEDs). As part of the NEMESIS project, we have performed a large-scale SED fitting analysis of the Orion star formation complex (OSFC).}  
    {We aim to derive key physical parameters, including temperature, luminosity, mass, and age, for a large sample of sources in the OSFC using synthetic SED models. Our goal is to establish a statistically robust characterization of the stellar population and its evolutionary state across the entire complex.}  
    {We utilize a set of new radiative transfer model SEDs that span a variety of geometries and parameter spaces. These SEDs are fitted to multi-wavelength photometric data from optical to submillimeter wavelengths. We conducted SED fitting on a sample of 15,396 sources. Among these, 5,062 have at least a reliable W3 (12 $\mu$m) detection at longer wavelengths, and 63 have sub-millimeter detections in APEX/SABOCA at 350 $\mu$m or APEX/LABOCA at 870 $\mu$m. The resulting physical parameters are cross-referenced with stellar evolutionary tracks to ensure consistency with theoretical predictions.}  
    {The derived parameters reveal a diverse stellar population. Sources placed on the Hertzsprung-Russell diagram show distinct evolutionary sequences. The results are provided with varying levels of completeness and reliability, depending on the available data for each source. The catalog includes quality indicators such as the flux code, which represents the longest detected wavelength for each source, as well as Prob\_W3 and Prob\_W4 values that quantify the reliability of the AllWISE W3 and W4 detections. All results, including SED fitting outcomes, uncertainty estimates, and source metadata, are publicly available in a comprehensive CDS table.}  
    {This dataset provides a statistically significant view of the evolutionary processes within the OSFC. The publicly accessible dataset offers a valuable resource for future studies on star and planet formation.}

   \keywords{catalogs --
                Stars: formation --
                Accretion, accretion disks
               }

   \maketitle


\section{Introduction}
Star formation occurs in molecular clouds when a rotating, magnetized cloud collapses under its gravity. Despite this relatively simple description, the physical processes that govern star formation are still a significant area of active research in astronomy. Young stellar objects (YSOs) represent an early phase in stellar evolution, often associated with the surrounding gas and dust from which they form. These objects are frequently observed in regions of active star formation, such as stellar nurseries and molecular clouds. Among these regions, the Orion Star Formation Complex (OSFC), located at a distance of approximately 400 parsecs \citep{menten07,oriondist18, kuhn19}, is the closest major star-forming region and one of the most extensively studied. It provides critical insights into both low- and high-mass star formation, the evolution and dispersal of star-forming clouds, and the dynamics of the interstellar medium.

Research on YSOs has unveiled a variety of complex detection and characterization methods across different scientific domains. Spectroscopic techniques, exemplified by \cite{briceno19}, identified over 2000 T Tauri stars in Orion OB1 by analyzing the Li I line absorption and H$\alpha$ emission characteristics. Further expanding our understanding, \cite{kounkel18} conducted a kinematic analysis, detecting approximately 9000 kinematic members using \textit{Gaia} DR2 \citep{Gaiadr2} positions, proper motions, and radial velocity data from the Apache Point Observatory Galactic Evolution Experiment \citep[APOGEE,][]{apogee17}. Additionally, photometric variability research has uncovered multiple mechanisms influencing stellar observation, such as stellar rotation, magnetic spots, disk dust dimming, and fluctuations in the accretion rate \citep{ysovar11, cody10, cody14, cody18}.

One of the key tools for understanding the evolutionary stages of YSOs is the spectral energy distribution (SED), which represents an object's energy output across a wide range of the electromagnetic spectrum, from gamma rays to radio waves. By plotting flux as a function of wavelength, the shape of the SED reveals crucial information about the physical properties and evolutionary stages of YSOs. In their early stages, YSOs exhibit a significant infrared excess, primarily due to circumstellar disks and envelopes that reprocess stellar radiation into thermal emission. As these objects evolve, their SEDs gradually transition from being dominated by circumstellar material to resembling stellar photospheres, with a diminishing infrared excess. This evolution reflects the dissipation of circumstellar material as the star approaches the main sequence. Thus, analyzing SED shapes is vital for understanding both the formation and early evolution of stars and planetary systems.

YSOs are often classified into different evolutionary classes based on their infrared SEDs, with a spectral index $\alpha$ serving as a key parameter to differentiate between classes \citep{lada87, adams87, andre94, evans09, dunham14}. The infrared spectral index $\alpha$ measures how the flux density, F$_\lambda$, changes with wavelength, $\lambda$, and is calculated using the equation:

\[
\alpha = \frac{\mathrm{d} \log (\lambda F_\lambda)}{\mathrm{d} \log \lambda}.
\]

In practice, $\alpha$ is computed by performing a linear fit to the SED in logarithmic space, allowing researchers to infer the evolutionary phases of YSOs based on their infrared emission characteristics.

Initially, three classes (Class I, II, III) were defined \citep{lada1984,lada87}, and later expanded to five \citep[Class 0, I, Flat, II, III,][]{greene94}. Class 0 represents the earliest collapse phase, dominated by an envelope \citep{andre1993}. Classes I, II, and III correspond to protostars, pre-main-sequence stars with circumstellar disks, and evolved pre-main-sequence stars or main-sequence stars, respectively. Flat-spectrum sources, which lie between Classes I and II, have an ambiguous evolutionary status due to geometric effects, extinction, and observational biases \citep{whitney03a, whitney03b, whitney13, robitaille2006, crapsi08}.

A common technique for inferring the physical properties of YSOs is SED fitting, which compares observed SEDs with that of theoretical models to derive parameters such as temperature, luminosity, and characteristics of the circumstellar matter. While this method may seem straightforward, accurately determining the properties of YSOs from SED models is challenging due to model limitations and observational uncertainties. Traditional model grids are often biased due to specific theoretical assumptions or limited parameter sampling. The models introduced by \citet[][hereafter R17]{robitaille2017} offer broader geometric diversity and random parameter sampling, enhancing their applicability. However, these models lack certain critical parameters, such as core mass, and can sometimes produce unphysical scenarios. To address these limitations, \citet[][hereafter R24]{richardson24} recently updated the \citetalias{robitaille2017} models to improve their physical realism and expand their utility.

In constructing comprehensive SEDs, flux measurements across a wide range of wavelengths are essential. For YSOs, the infrared excess -- arising from heated dust -- provides crucial information about circumstellar environments, but it can obscure the underlying stellar emission. As a result, accurate SED fitting is critical for disentangling the contributions of the star and its surrounding material.

The Novel Evolutionary Model for the Early Stages of Stars with Intelligent Systems (NEMESIS) \footnote{\url{https://nemesis.univie.ac.at}}
project aims to revolutionize our understanding of star formation by leveraging artificial intelligence to analyze the largest, panchromatic YSO dataset \citep[see, e.g.,][]{roquette2024, marton24, hernandez2024, madarasz2024}. In addition to compiling the largest YSO catalog for the OSFC, NEMESIS enhances data quality through advanced machine learning and deep learning techniques \citep{marton24, madarasz2024}. As part of the  NEMESIS project, we have undertaken an extensive review of existing catalogs and literature to compile the most comprehensive YSO catalog for the OSFC, including a detailed review of data from the Wide-field Infrared Survey Explorer \citep[\textit{WISE,}][]{wright10} and the \textit{Herschel} Space Observatory \citep{Herschel10}. The OSFC covers a field of view spanning 564 square degrees, between  74.2°<RA<–92° and  $-14.1$°<Dec<17.6°.

Our work includes 27,879 sources with detailed positional information, flux measurements, and associated uncertainties across multiple wavelengths \citep{roquette2024}, and it represents the largest data set to date to which SED fitting has been applied for the classification of YSOs and the derivation of stellar parameters. We employ the updated radiative transfer model SEDs from \citetalias{robitaille2017} and \citetalias{richardson24}.

This paper is organized as follows: Section~\ref{Data} provides an overview of the surveys and catalogs used in our analysis, along with statistical details on the photometric data. Section~\ref{methods} outlines the updated \citetalias{richardson24} model, and describes the SED fitting methodology. The results of the SED fitting analysis are presented in Section~\ref{results}. Section~\ref{discussion} discusses the findings, and Section~\ref{conclusion} summarizes our conclusions.


\section{Data}
\label{Data}
Our SED analysis incorporates data from multiple astronomical surveys and instruments, including \textit{Gaia} DR3 \citep{Gaiadr3}, 2MASS \citep{skrutskie06}, \textit{WISE} \citep{wright10}, \textit{Spitzer}/IRAC \citep{fazio04}, \textit{Spitzer}/MIPS \citep{rieke04}, \textit{Herschel} \citep{Herschel10}, and the Atacama Pathfinder Experiment (APEX) \citep{stutz13}. Below, we summarize the contributions of each dataset and its relevance to this study.

\subsection{\textit{Gaia} DR3}
\textit{Gaia} is one of the primary datasets utilized in our YSO catalog, providing optical photometric and astrometric data for the largest number of YSOs. \textit{Gaia} conducts a full-sky survey, delivering observations in three photometric bands: $G$ (0.58~$\mu$m), $BP$ (0.50~$\mu$m), and $RP$ (0.76~$\mu$m). Approximately 20,000 sources in our catalog have available \textit{Gaia} photometry in all three bands.

In our OSFC catalog we used parallax data from the third data release of the \textit{Gaia} mission \citep[\textit{Gaia} DR3,][]{Gaiadr3} to determine accurate distances. To ensure reliability, we applied a 3-sigma cutoff to filter out sources with uncertain parallax data, thus reducing the impact of measurement uncertainties and excluding outliers. This process identified 22,275 sources (out of 27,879) with reliable parallax measurements. Of these, approximately 17,500 sources fall within the distance range of 200 to 600 parsecs \citep{kounkel18, grossschedl2019}, which approximately corresponds to the OSFC. For these 17,500 sources, we utilized the precise \textit{Gaia} distance in further SED fitting. For the remaining sources, where \textit{Gaia}’s distance measurements were less reliable, unavailable, or outside the 200 to 600 parsecs range, we continued SED fitting using an average OSFC distance of 400 parsecs.

Despite its high precision, \textit{Gaia}'s distance estimates in star-forming regions can be unreliable. Factors such as interstellar dust, gas, and the diffuse environment of OSFC can affect parallax accuracy. Additionally, YSOs are often surrounded by dense envelopes or disks, which obscure optical observations. Variability caused by dynamic accretion processes further complicates these measurements. A similar challenge has been discussed in the context of AGB stars by \cite{andriantsaralaza22}. Given these limitations, we chose to retain all sources in our catalog, including those with less reliable or absent distance measurements, rather than imposing stricter selection criteria. The methodology used for data compilation is detailed in \cite{roquette2024}. Further discussion of the SED fitting process is provided in Section \ref{sed_fitting}.

\subsection{2MASS}
The Two Micron All-Sky Survey (2MASS) was a ground-based survey that mapped the entire sky in near-infrared wavelengths in the J (1.235 $\mu$m), H (1.662 $\mu$m), and K$_\mathrm{s}$ (2.159 $\mu$m) bands. The 2MASS Point Source Catalog \citep{skrutskie06} provides near-photometry for the majority, over 23,000 sources of our OSFC YSOs catalog.

\subsection{\textit{Spitzer} Space Telescope}
Up until the launch of the \textit{James Webb Space Telescope} (JWST) \citep{JWST06}, \textit{Spitzer} Space Telescope \citep{Spitzer04} was the most sensitive infrared telescope. While JWST is a highly advanced and extremely sensitive observatory, it is still relatively new, and its primary focus on targeted studies rather than large-scale surveys means that \textit{Spitzer} continues to provide the most sensitive infrared data for a greater number of sources. The Spitzer enabled extensive exploration of YSOs within the Milky Way through its Infrared Array Camera (IRAC) and Multiband Imaging Photometer (MIPS). IRAC observed at 3.6, 4.5, 5.8, and 8.0 $\mu$m, while MIPS covered 24 $\mu$m (MIPS1) and 70 $\mu$m (MIPS2). These comprehensive surveys, including the Galactic Legacy Infrared Midplane Survey Extraordinaire \citep[GLIMPSE][]{benjamin03}) and the MIPS Galaxy (MIPSGAL) surveys have played a crucial role in examining numerous star-forming regions \citep[e.g.,][]{allen04, megeath04, megeath05, gutermuth09, megeath12, povich2011, dunham2015}. In our OSFC catalog, approximately 10,000 sources have IRAC photometry, while 2,500 sources have data at 24 $\mu$m from MIPS. No sources in the catalog have photometry at 70 $\mu$m.

\subsection{\textit{WISE}}
While \textit{Spitzer} provides sensitive infrared data for a greater number of sources, it still does not offer full-sky coverage. In this context, the \textit{WISE} survey \citep{wright10} is crucial due to its comprehensive full-sky coverage. As \textit{WISE} is the only full-sky near-to-mid infrared survey, it often provides the primary mid-infrared photometry for sources. For \textit{WISE} W1 and \textit{WISE} W2 data we used the Cat\textit{WISE}2020 Catalog \citep{eisenhardt20, marocco21}. The Cat\textit{WISE}2020 Catalog is a comprehensive dataset containing positions, brightnesses, and motion measurements for over 1.89 billion sources, derived from \textit{WISE} and NEO\textit{WISE} \citep{mainzer14} observations spanning from 2010 to 2018. It focuses on wavelengths at 3.4 $\mu$m (W1) and 4.6$\mu$m (W2), making it valuable for studying stars, galaxies, and other celestial objects. Cat\textit{WISE}2020 uses 8 years of \textit{WISE} and NEO\textit{WISE} data, compared to All\textit{WISE}, which is limited to 4 years. Additionally, Cat\textit{WISE}2020 employs the ``crowdsource'' detection algorithm \citep{schlafly19} which is more effective in dense regions, such as the Galactic plane, where source blending is common.

The \textit{WISE} mission was designed primarily for studying infrared-bright galaxies, brown dwarfs, and near-Earth asteroids. Consequently, detecting YSOs or evolved stars embedded in diffuse environments can be problematic, and photometry, particularly in the \textit{WISE} W3 and W4 bands (at 12 and 22 $\mu$m), can be less reliable. This issue has been explored in various studies \citep[e.g.,][]{koenig14, marton19, silverberg18}, which indicated that previous methods were overly aggressive in filtering out spurious \textit{WISE} W3 and W4 band photometry. To address this issue, we have updated these methodologies and incorporated new machine-learning techniques. Details of these improvements and the revised All\textit{WISE} catalog are presented in \cite{roquette2024}. The updated All\textit{WISE} catalog now includes probabilities indicating the likelihood that W3 and W4 band detections are real. For the sources in the OSFC, we provide the reliability values labeled as \textit{Prob\_W3} and \textit{Prob\_W4} enhancing the robustness of the SED fitting analysis.

\subsection{\textit{Herschel} Space Observatory}
\textit{Herschel} Space Telescope \cite{Herschel10} is another major cornerstone in infrared astronomy. With a 3.5-meter primary mirror, \textit{Herschel} was a space telescope operated as an observatory and performed diverse observing programs aimed at varied scientific objectives. Therefore, it didn't cover the full sky and only $\sim$8\% of the sky was observed with \textit{Herschel}. The \textit{Herschel} Space Observatory, which conducted many scientific observation programs, provided a catalog of point sources \citep{marton17} at 70, 100, and 160 $\mu$m. As part of \href{https://nemesis.univie.ac.at/}{NEMESIS}, this catalog was improved \citep{marton24, madarasz2024} using advanced deep learning methods to ensure high reliability and to filter out unreliable sources. The final updated catalog, recently published by \cite{marton24}, provides reliable far-infrared (FIR) photometry for thousands of objects. However, since \textit{Herschel} was not a full-sky survey, its coverage is patchy, and only hundreds of YSOs in the OSFC have \textit{Herschel} photometry.

\subsection{APEX}
To extend the SEDs into the sub-millimeter range, we have included data from the Atacama Pathfinder EXperiment \citep[APEX,][]{stutz13}. APEX is a submillimeter telescope located on the Chajnantor Plateau in the Atacama Desert, Chile. Operating at a high altitude of 5,100 meters, it is optimized for observing the universe at millimeter and submillimeter wavelengths, crucial for studying cold gas and dust in space, star formation, and distant galaxies. LABOCA (Large APEX Bolometer Camera) is one of the primary instruments on APEX. It operates at 870 $\mu$m (350 GHz), in the submillimeter range. SABOCA (Submillimeter APEX Bolometer Camera) is another bolometer camera on APEX.
It operates at a shorter wavelength of 350 $\mu$m (850 GHz), compared to LABOCA. For a few hundred sources, we have APEX/SABOCA data at 350$\mu$m and APEX/LABOCA data at 870 $\mu$m available.

\subsection{Photometric Data Compilation and Data Quality for SED Fitting}
We employed a total of 20 photometric filters for our study. The final dataset used for SED fitting includes flux measurements and associated errors from \textit{Gaia}, 2MASS, IRAC, Cat\textit{WISE}, All\textit{WISE}, PACS, and APEX. A notable addition to this table is the \textit{flux\_column}, which indicates the farthest detected flux (non-NaN value) for each source. This column was specifically created to provide insight into the longest-wavelength detection for each source, helping to better understand their observational properties. Each code corresponds to a specific filter, as detailed below. The flux codes follow a hierarchical naming system, where shorter wavelengths are labeled first (e.g., I1 for IRAC 3.6\,$\mu$m), and longer wavelengths are progressively labeled according to the farthest detected flux.

\begin{table}[h]
    \centering
    \caption{Flux Code Abbreviations}
    \label{flux_codes}
    \begin{tabular}{ll}
        \hline
        \hline
        Flux Code & Longest Detected Wavelength \\
        \hline
        I1 & \textit{Spitzer}/IRAC 3.6\,$\mu$m \\
        I2 & \textit{Spitzer}/IRAC 4.5\,$\mu$m \\
        I3 & \textit{Spitzer}/IRAC 5.8\,$\mu$m \\
        I4 & \textit{Spitzer}/IRAC 8.0\,$\mu$m \\
        W1 & \textit{AllWISE}/W1 3.4\,$\mu$m \\
        W2 & \textit{AllWISE}/W2 4.6\,$\mu$m \\
        W3 & \textit{AllWISE}/W3 12\,$\mu$m \\
        W4 & \textit{AllWISE}/W4 22\,$\mu$m \\
        M1 & \textit{Spitzer}/MIPS 24\,$\mu$m \\
        P1 & \textit{Herschel}/PACS 70\,$\mu$m \\
        P2 & \textit{Herschel}/PACS 100\,$\mu$m \\
        P3 & \textit{Herschel}/PACS 160\,$\mu$m \\
        S  & \textit{APEX}/SABOCA 350\,$\mu$m \\
        L  & \textit{APEX}/LABOCA 870\,$\mu$m \\
        \hline
    \end{tabular}
\end{table}

\begin{figure*} 
\centering
\includegraphics[scale=0.50]{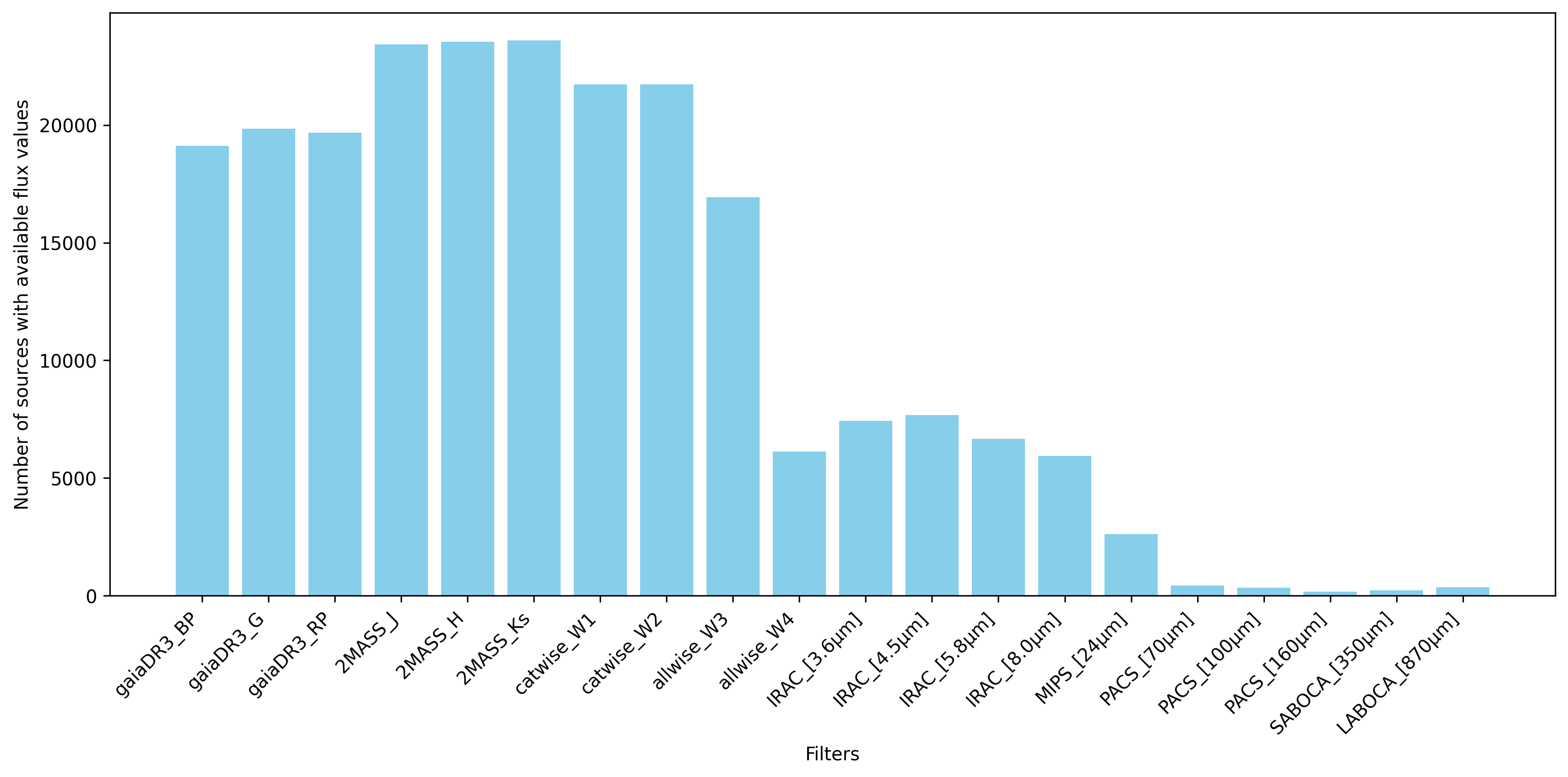}
\caption{Available flux counts for each filter.}
\label{flux_histogram}
\end{figure*}

Additionally, we introduced the \textit{Prob\_W3} and \textit{Prob\_W4} columns to assess the reliability of All\textit{WISE} W3 and W4 photometry, enabling us to verify whether these detections are genuine. Figure~\ref{flux_histogram} displays the number of sources with available photometric data for each filter. We performed the SED fitting analysis for sources with at least four valid data points. Out of the total 27,879 sources, 24,747 meet this criterion, ensuring a robust dataset for the fitting process.

\subsection{Multiplicity Considerations}

The multiplicity labels in our study are based on the main Orion YSO catalog compiled by \citet{roquette2024}, which provides a comprehensive assessment of binarity and source contamination. The catalog underwent rigorous cleaning procedures, incorporating literature-based searches and advanced techniques, including machine learning and deep learning methods, to identify and flag potential binaries. As a result, the final dataset used in our SED fitting analysis was already scientifically curated to exclude known binaries and ambiguous sources, ensuring a more reliable interpretation of the fitted models. However, we acknowledge that observational limitations remain, and some unresolved binaries may still be present in the catalog.

\section{Methods}\label{methods}
\subsection{SED Models}\label{R24_model}
One of the most effective methods for characterizing YSOs, especially when they are unresolved, is by fitting their SEDs to pre-computed grids of radiative transfer models. Various SED model grids have been developed for this purpose, including those by \cite{robitaille2007}, \cite{furlan16}, \cite{haworth18} and \cite{zhang18}. However, these models often assume specific star formation theories, limiting their applicability for testing different star formation scenarios. To address these limitations, the SED models presented by \citetalias{robitaille2017} offer a comprehensive set of templates that are agnostic to accretion history and stellar evolution models. These models span various geometries and parameter spaces, making them versatile tools for determining YSO properties. Nonetheless, \citetalias{robitaille2017} models had certain limitations, such as the absence of core mass values and the inclusion of unrealistic physical models.

Recently, the \citetalias{robitaille2017} models have undergone significant enhancements by \citetalias{richardson24}, improving their utility and accuracy in determining YSO properties. The SEDs have been convolved with filters from JWST \citep{JWST06}, Paranal \citep{paranal_vista15}, additional \textit{Herschel} filters, and Atacama Large Millimeter Array (ALMA) \citep{ALMA09} bands 3 and 6. The infrared spectral index \(\alpha\) has been calculated for all SEDs within each aperture, facilitating the classification of YSOs into observational classes based on their evolutionary states. 

\citetalias{richardson24} determined the spectral index $\alpha$ as the slope of the line in log-space connecting the flux measurements at 2 and 25 $\mu$m and defined the classes as below.

\begin{itemize}
    \item \textbf{Class I}: \(\alpha \geq 0.3\)
    \item \textbf{Flat}: \(-0.3 \leq \alpha < 0.3\)
    \item \textbf{Class II}: \(-1.6 \leq \alpha < -0.3\)
    \item \textbf{Class III}: \(\alpha < -1.6\)
    \label{spectral_index}
\end{itemize}

The \citetalias{robitaille2017} model, along with its updated version by \citetalias{richardson24}, consists of multiple components, including the star, disk, infalling envelope, bipolar cavities, and ambient medium. These components are modeled with varying levels of complexity, ranging from as few as two parameters in the simplest models to as many as twelve in the most detailed configurations. The models span a wide range of evolutionary stages, from the youngest, most deeply embedded YSOs to pre-main-sequence/main-sequence stars with minimal or no surrounding disk. Table \ref{model_sets} summarizes the available model sets in both \citetalias{robitaille2017} and the updated \citetalias{richardson24} model. The updated model explicitly includes the luminosity of the central source for each configuration, enabling checks of the model's position on the Hertzsprung-Russell (H-R) diagram. In contrast to previous versions, where outer envelope radii were not included and inner radii were either expressed in terms of the variable \(R_\mathrm{sub}\) or omitted, the updated model provides these parameters explicitly. Additionally, the updated model offers further quantities, such as core mass, average dust temperature, and circumstellar extinction
\(A_\mathrm{v}\), facilitating a more comprehensive characterization of YSOs.

\citetalias{richardson24} also adopts a \textit{stagification} scheme to classify YSOs, utilizing the definitions from \cite{crapsi08} and \cite{evans2009} as a foundation. This scheme categorizes YSOs into four distinct stages based on the characteristics of their envelopes and central temperatures:

\begin{itemize}
    \item \textbf{Stage 0}: The envelope mass ($M_{\text{env}}$) is greater than 0.1 $M_{\odot}$, and the stellar temperature ($T_{\star}$) is less than 3000 K.
    \item \textbf{Stage I}: The envelope mass ($M_{\text{env}}$) remains greater than 0.1 $M_{\odot}$, but the stellar temperature ($T_{\star}$) exceeds 3000 K.
    \item \textbf{Stage II}: The envelope mass ($M_{\text{env}}$) is less than 0.1 $M_{\odot}$, and a disk is present.
    \item \textbf{Stage III}: The object is a bare pre-main-sequence star, with neither an envelope nor a disk.
\end{itemize}

Class is used for the observed SED classification, while Stage refers to the physical configuration. This classification/stagification system provides a structured framework for analyzing YSOs, offering clear distinctions between different evolutionary stages based on observable physical properties.

\subsection{SED Fitting}\label{sed_fitting}
SED fitting involves comparing the observed SEDs of YSOs to a set of theoretical models that account for different physical components, such as the star, disk, infalling envelope, bipolar cavities, and ambient medium. 

In our study of YSOs, we fit their SEDs using newly updated radiative transfer model SEDs from \citetalias{richardson24}, utilizing the \citetalias{robitaille2017} SED fitting tool. Details of the \citetalias{richardson24} model are provided in the previous Section~\ref{R24_model}. In addition to several convolved filters of the \citetalias{richardson24} model, we have also added \textit{Gaia} filters to the model. We have convolved the SEDs with filters from \textit{Gaia} DR3 G, BP, and RP. The filter transmission curves were obtained from the SVO's Filter Profile Service \citep{rodrigo12, rodrigo20}.

As shown in Fig.~\ref{flux_histogram}, photometric data are not available for all sources across the 20 different bands. For our SED fitting analysis, we include all sources with at least four valid data points, regardless of whether they have detections at longer wavelengths. This inclusive approach is designed to minimize selection bias. Restricting sources to those with detections at longer wavelengths could exclude more evolved objects, which may no longer be detectable in these wavelength ranges. 

SED fitting has been conducted for the 24,747 sources (see Section~\ref{Data}). For most of the sources, we used the optical \textit{Gaia} G, BP and RP, 2MASS near-infrared (J, H, and Ks) photometry with mid-infrared photometry from either \textit{Spitzer}/IRAC (3.6, 4.5, 5.8, and 8.0 $\mu$m) or/and \textit{WISE} W1, W2 and W3 (3.4, 4.6, 12.0 $\mu$m). For a few thousand sources, we have been able to include \textit{Spitzer}/MIPS at 24 $\mu$m or/and \textit{WISE} W4 photometry at 22 $\mu$m. Only several hundred sources have far-infrared photometry from \textit{Herschel} PACS and sub-millimeter data from APEX, and we have constructed SEDs for these sources ranging from optical to far-infrared (0.5 to 870 $\mu$m). 

For sources with reliable \textit{Gaia} distance measurements, we divided them into distance bins and performed individual fittings for each bin, allowing a 10\% variation from the median distance. For sources where \textit{Gaia}’s distance measurements were unreliable, unavailable, or fell outside the 200 to 600 parsecs range, we used a mean distance of 400 parsecs with a 10\% variation. 

We utilized the updated Galactic dust extinction estimates from \cite{SandF11}, which were derived using the IRSA/IPAC Galactic Dust Map webpage\footnote{\url{https://irsa.ipac.caltech.edu/applications/DUST/}} in our analysis. Our findings indicate that 95\% of the sources have extinction values ranging from 0 to 50 magnitudes. We accounted for interstellar extinction (\(A_\mathrm{V}\)) in the range of 0 to 50 mag, in addition to considering the extra extinction from the circumstellar environment of the YSOs. For this purpose, we applied the extinction law from \cite{weingartner01} with a value of \(R_\mathrm{V} = 4.0\).

To evaluate model fits, we used a Bayesian approach that assigns relative probabilities to models of varying complexity. We employed a threshold parameter (\textbf{F=3}) to identify “good” fits. This selection follows the SED fitter model selection syntax, where ‘F’ indicates that we retain all models satisfying the condition that the reduced difference from the best-fit model, defined as $\chi^2-\chi^2_{\rm best}$ per datapoint, is below the chosen threshold value of 3. This ensures that we consider models that provide a statistically reasonable fit while excluding those with significantly higher deviations from the best-fit solution. This method helps avoid overfitting, particularly for sources with limited data points.

\section{Results}
\label{results}
\begin{figure}
\centering
\includegraphics[scale=0.40]{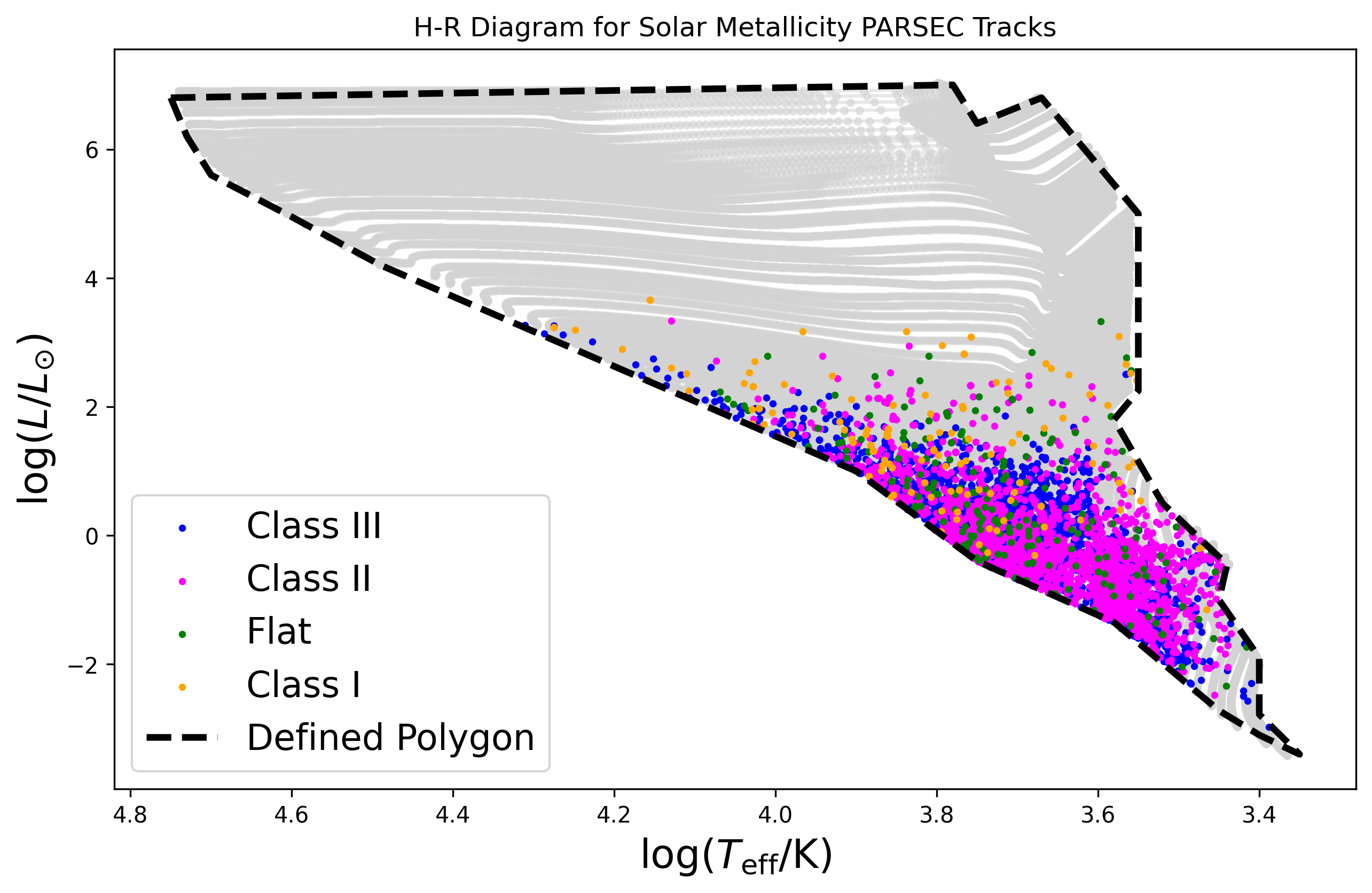}\
\caption{Luminosity ($L$) as a function of effective temperature ($T_\mathrm{eff}$). Values are derived from the SED fitting of 15,396 best-fit models and overlaid on the PARSEC evolutionary tracks (with solar metallicity). The data points are color-coded to represent different stellar classes.}
\label{HR_diagram}
\end{figure}

We performed SED fitting for a total of 24,747 sources. The updated model set was constructed to provide uniform coverage of the parameter space; however, some models remain non-physical. To ensure the inclusion of only physically plausible models, we compared their positions on the H-R diagram with PAdova and TRieste Stellar Evolution Code \citep[PARSEC][]{bressan12, chen14, chen15, tang14} evolutionary tracks. Based on this selection criterion, we successfully identified best-fit models for 15,396 sources (see Fig.~\ref{HR_diagram}). Table~\ref{model_sets} summarizes the model sets and their components, as described in \citetalias{robitaille2017}, along with the number of sources matched to each set.

In Fig.~\ref{HR_diagram}, we present the temperature and luminosity of each source derived from the best-fit parameters on the H-R diagram, overlaid with evolutionary models from the PARSEC tracks. The PARSEC evolutionary tracks are theoretical models describing stellar evolution, computed using the PARSEC code, a widely adopted tool in stellar astrophysics. This release introduces a new implementation of pre-main sequence (PMS) evolution. During this phase, the star is fully convective, homogeneously mixed, and lacks active nuclear reactions. From this initial configuration, the star evolves at constant mass, progressing through the PMS contraction phase and subsequent stages of stellar evolution. Consequently, these tracks typically begin before the Zero-Age Main Sequence (ZAMS) for most stellar masses, spanning a range of 0.1 to 350 M$_\odot$ \citep{bressan12, chen15}. The ZAMS marks the point where stellar models traditionally begin, as the physics in this phase is more stable and simpler compared to the complex processes of the PMS phase. Since our OSFC catalog contains sources younger than the ZAMS, it is particularly advantageous to utilize the PARSEC tracks for our analysis.

Among the approximately 15,400 best-fit models, about 8,950 correspond to sources with All\textit{WISE} W3 detections. However, as noted earlier, All\textit{WISE} W3 and W4 detections are not always reliable. To address this, we present a subset of more reliable SED fitting results for sources with robust detections. Specifically, we selected sources with a W3 (12~$\mu$m) detection at longer wavelengths and a detection probability (\textit{Prob\_W3}) greater than 50\%. This criterion identifies a total of 5,062 sources, referred to as the high-confidence subset. 

Using the PARSEC evolutionary tracks, we estimated the mass and age of approximately 15,400 best-fit models. Stellar masses were determined by identifying the nearest evolutionary track, while stellar ages were interpolated from the closest point on the track corresponding to each source's position in the H-R diagram.

For approximately 15,400 sources with successful fits, the complete results are available in the CDS table. This comprehensive dataset includes crucial SED fitting information such as \textit{flux\_code} indicating the longest wavelength detection for each source, along with the reliability probabilities for the W3 and W4 photometry, \textit{Prob\_W3} and \textit{Prob\_W4}, the number of valid data points used (ndpt), $\chi^2$ per data point for the best-fit model ($\chi^2$ per ndpt), and the number of fits (nfits) with $\chi^2$/ndpt between ($\chi^2$/ndpt)${set}$ and ($\chi^2$/ndpt)${set}$ + 2 within the most likely model set. Additionally, the table provides physical parameters estimated from the SED fitting—such as temperature and luminosity—alongside calculated values for Class, Stage, and spectral index.

\begin{table*}
    \centering
        \caption{Models as described by \citetalias{robitaille2017}, the number of sources fitted with the given model and components of the models.}
    \label{model_sets}
    \begin{tabular}{|l|c|l|}
        \hline
        Model & \# Best-Fit & Components \\ 
        \hline
        s---s-i   & 6487 & star \\ 
        sp--s-i   & 601  & star + passive disk; $R_{\text{inner}} = R_{\text{sub}}$ \\ 
        sp--h-i   & 650  & star + passive disk; variable $R_{\text{inner}}$ \\ 
        s---smi   & 5113 & star + medium; $R_{\text{inner}} = R_{\text{sub}}$ \\ 
        sp--smi   & 572  & star + passive disk + medium; $R_{\text{inner}} = R_{\text{sub}}$ \\ 
        sp--hmi   & 661  & star + passive disk + medium; variable $R_{\text{inner}}$ \\ 
        s-p-smi   & 19   & star + power-law envelope + medium; $R_{\text{inner}} = R_{\text{sub}}$ \\ 
        s-p-hmi   & 46   & star + power-law envelope + medium; variable $R_{\text{inner}}$ \\ 
        s-pbsmi   & 134   & star + power-law envelope + cavity + medium; $R_{\text{inner}} = R_{\text{sub}}$ \\ 
        s-pbhmi   & 75   & star + power-law envelope + cavity + medium; variable $R_{\text{inner}}$ \\ 
        s-u-smi   & 148   & star + Ulrich envelope + medium; $R_{\text{inner}} = R_{\text{sub}}$ \\ 
        s-u-hmi   & 100   & star + Ulrich envelope + medium; variable $R_{\text{inner}}$ \\ 
        s-ubsmi   & 177   & star + Ulrich envelope + cavity + medium; $R_{\text{inner}} = R_{\text{sub}}$ \\ 
        s-ubhmi   & 161   & star + Ulrich envelope + cavity + medium; variable $R_{\text{inner}}$ \\ 
        spu-smi   & 98   & star + passive disk + Ulrich envelope + medium; $R_{\text{inner}} = R_{\text{sub}}$ \\ 
        spu-hmi   & 109   & star + passive disk + Ulrich envelope + medium; variable $R_{\text{inner}}$ \\ 
        spubsmi   & 94   & star + passive disk + Ulrich envelope + cavity + medium; $R_{\text{inner}} = R_{\text{sub}}$ \\ 
        spubhmi   & 151   & star + passive disk + Ulrich envelope + cavity + medium; variable $R_{\text{inner}}$ \\ 
        \hline
    \end{tabular}
    \tablefoot{The model set names, as described by \citetalias{robitaille2017}, with the number of sources identified in each set are presented in the first and second columns, respectively. The model set names consist of seven characters, each indicating the presence of a specific component. The third column details the presence of specific components in the following order: \textbf{s} for the star, \textbf{p} for the passive disk, \textbf{p} or \textbf{u} for power-law or Ulrich envelope, \textbf{b} for bipolar cavities, \textbf{h} for the inner hole, \textbf{m} for ambient medium, and \textbf{i} for interstellar dust. A dash ($-$) indicates the absence of a component. R$_{inner}$ refers to the inner radius of the disk, envelope, or ambient medium, depending on the components included, while R$_{sub}$ represents the dust sublimation radius.}
\end{table*}

We found that best-fit models incorporating Gaia data are relatively few, and their inclusion does not significantly improve the best-fit determination. In cases where Gaia data is available, the fitting process tends to favor models with circumstellar material (e.g., disks or envelopes), even for sources lacking detections beyond W3 (12~$\mu$m). This effect arises because Gaia strongly constrains the optical brightness, whereas no equivalent constraint exists in the infrared. Consequently, the fitter may interpret any missing flux as being absorbed by circumstellar dust rather than interstellar extinction. Furthermore, chi-squared values were consistently higher when Gaia data was used, indicating that its inclusion does not always lead to a statistically better fit.

The table~\ref{selected_rows} in Appendix~\ref{SED_18} presents the SED fitting results for a sample of 20 YSOs. The corresponding SED plots are also presented in Fig~\ref{SEDs} in Appendix~\ref{SED_18}. The table highlights key parameters used in SED fitting, providing critical insights into the reliability and limitations of the fitting process. The inclusion of \textit{ndpt} highlights the diversity in data completeness across the sample, which directly impacts the robustness of the SED fitting. Sources with a higher number of data points provide more constrained fits, whereas sources with fewer data points may result in greater uncertainties. Similarly, the \textit{flux\_code} and detection probabilities (\textit{Prob\_W3}, \textit{Prob\_W4}) serve as indicators of the quality and reliability of individual photometric measurements. For instance, higher probabilities signify confident detections, enhancing the reliability of flux contributions at those wavelengths. Conversely, lower probabilities or missing values introduce additional uncertainties in the model-fitting process. Accurate distances are critical for determining the intrinsic physical properties of YSOs. For sources without precise parallax measurements, we adopt the average distance to the OSFC (400~pc). While this assumption is necessary for some sources, it introduces a degree of systematic uncertainty in the derived parameters. Together, these data for the 20 selected sources provide essential context for understanding the reliability and limitations of the SED fitting, enabling a more informed interpretation of the results.

\section{Discussion}\label{discussion}
\subsection{Comparison with the results of \cite{kounkel18}}
\cite{kounkel18} provided key stellar properties, including effective temperature (\textit{T$_{\mathrm{eff}}$}), surface gravity (\(\log g\)), and radial velocity (RV), derived from the APOGEE spectra \citep{apogee17} for approximately 9000 stars in the OSFC. Among the 15,396 successful best fits from our analysis, 6,551 sources match those in the list provided by \cite{kounkel18}. Furthermore, when we focus on the SED fitting results of high-confidence sources, 3,482 of them match the list from \cite{kounkel18}.

\begin{figure}
\centering
\includegraphics[width=0.45\textwidth]{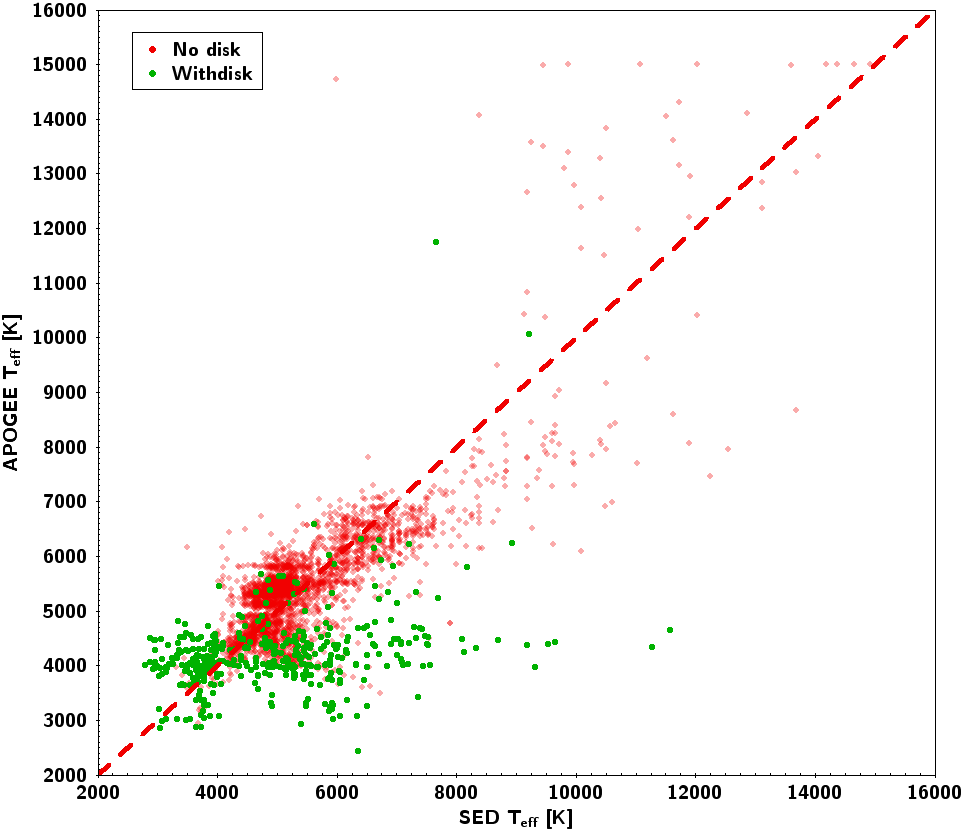}
\caption{Effective temperature (\(T_\mathrm{eff}\)) values from \cite{kounkel18} versus the \(T_\mathrm{eff}\) values derived from the SED fitting of 3,482 high-confidence sources. Sources, where the best fitting model includes a disk, are plotted with green dots.}
\label{teff}
\end{figure}

In Figure~\ref{teff}, we compare the effective temperature (\(T_\mathrm{eff}\)) values derived from the SED fitting of our high-confidence sources with those obtained by \citet{kounkel18} using APOGEE spectra. Despite the methodological differences in deriving these temperatures, the two sets of values show general agreement, and align well with those derived from APOGEE spectra, with a calculated Pearson's correlation coefficient of 0.76. However, we note that sources with a disk component (green points in Figure~\ref{teff}) tend to show larger scatter, with many appearing along a horizontal distribution. This is likely due to the presence of circumstellar material contributing significant infrared excess, which can introduce degeneracies in the SED-based temperature determination. In particular, embedded or accreting sources may have their stellar photosphere partially veiled by disk emission, leading to systematic differences in the inferred \(T_\mathrm{eff}\) compared to spectroscopic estimates.

To assess the impact of circumstellar disks, we computed the correlation coefficient separately for sources with and without a disk component. When excluding disk-bearing sources, Pearson's correlation coefficient improves to 0.82, suggesting that the SED-derived temperatures are more reliable in cases where the photospheric emission dominates.

\subsection{Color-Color diagrams}\label{ccd}

The identification and classification of YSOs in the OSFC have been extensively studied, particularly using color-color diagrams \citep[e.g.,][]{megeath04, megeath05, gutermuth09, megeath12, megeath16}. With the advent of \textit{Spitzer}, these classifications were further tested. Several color-color diagrams were defined to distinguish different classes of YSOs \citep{megeath04, megeath05, gutermuth09}, such as J$-$H vs. H$-$[4.5], and H$-$K vs. K$-$[4.5]. \cite{megeath04} and \cite{allen04} demonstrated that the IRAC color-color diagram ([3.6]-[4.5] vs. [5.8]-[8.0]) is a valuable tool for identifying young stars with infrared excess emission. Building on this, \cite{muzerolle04} combined MIPS 24 $\mu$m photometry with IRAC photometry and showed that the identified classes were almost entirely consistent with those identified using the [3.6]-[4.5] vs. [5.8]-[8.0] diagram, providing additional validation of this classification method.

\begin{figure*}
\centering

\includegraphics[scale=0.40]{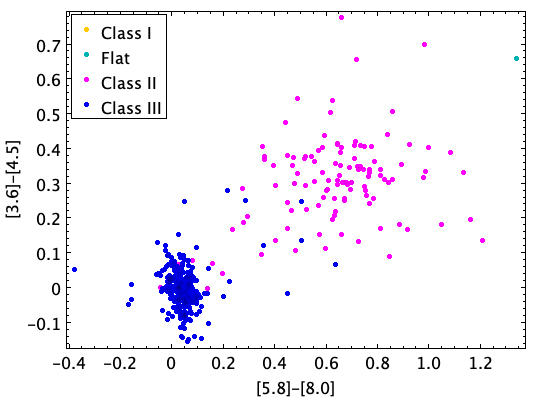}\
\includegraphics[scale=0.40]{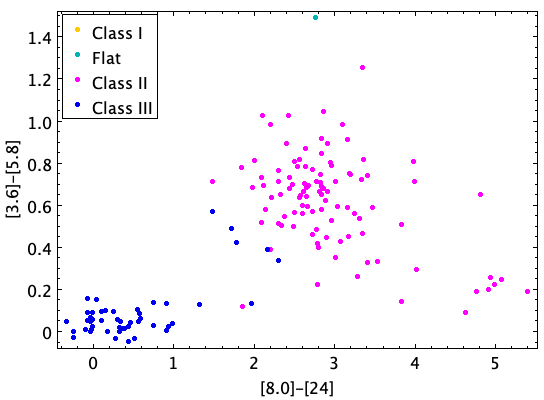}\
\includegraphics[scale=0.40]{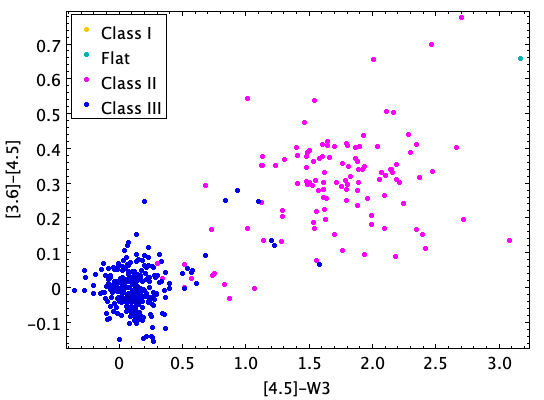}\
\includegraphics[scale=0.40]{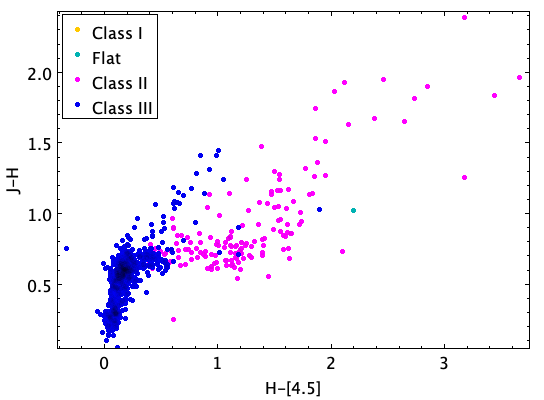}\
\includegraphics[scale=0.40]{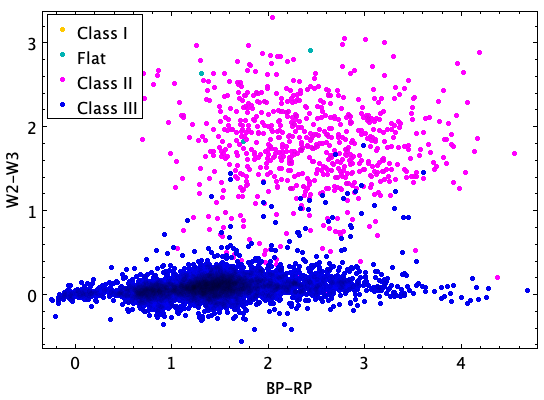}\
\includegraphics[scale=0.40]{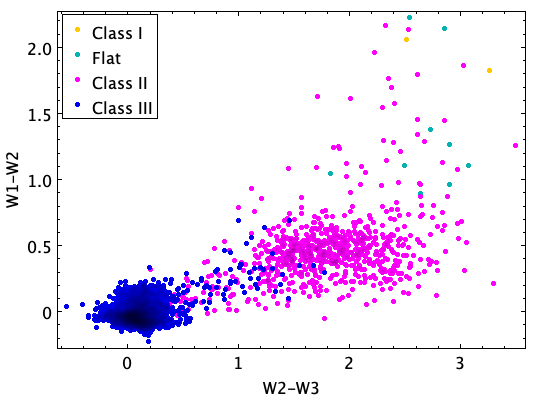}
\caption{Color-color diagrams for 5,062 high-confidence sources, with classes derived from SED fitting.}
\label{CCD}
\end{figure*}

Using \textit{Gaia} BP, G, RP, 2MASS J, H, Ks, \textit{Spitzer}/IRAC [3.6], [4.5], [5.8], [8.0], \textit{Spitzer}/MIPS [24], [70], and \textit{WISE} W1, W2, W3, and W4 magnitudes for our 5,062 high-confidence sources, we have generated all possible color-color diagrams. In Fig.~\ref{CCD} we present the most commonly used color-color diagrams for the 5,062 high-confidence sources. These sources have at least a W3 (12~$\mu$m) detection at longer wavelengths, with a detection probability (\textit{Prob\_W3}) greater than 50\%. This threshold ensures that the W3 detection is both robust and reliable. As illustrated in the color-color diagrams, the separation between classes is well-defined, further underscoring the reliability of the SED fitting results when based on a comprehensive and accurate dataset.

\subsection{Analysis of Classification Methods for YSOs}
\label{class_stage}

The updated \citetalias{richardson24} model introduces two complementary classification systems, as detailed in Section~\ref{R24_model}. The \textit{Class} designation is based on the shape of the SED, while the \textit{Stage} classification corresponds to the physical evolutionary phases of YSOs. Although the updated \citetalias{richardson24} model provides both class and stage definitions, we intentionally avoided comparing two systems derived from the same model. Instead, to perform an independent validation, we compared our SED-derived stages with classes determined by \citet{hernandez2024}.

\citet{hernandez2024} introduces a novel approach to YSO classification within the NEMESIS project, which integrates two-dimensional (2D) morphological data alongside traditional infrared spectral index (\(\alpha_\mathrm{IR}\)) classification. These classifications are derived by associating source morphologies obtained using Self-Organizing Maps (SOMs). As a baseline for this study, \citet{hernandez2024} adopts the infrared spectral index (\(\alpha_\mathrm{IR}\)) to classify YSOs based on the shape of their SEDs between \(2\text{--}24\,\mu\mathrm{m}\). This classification is performed using ordinary non-linear least-squares fitting.

The classification scheme used in \citet{hernandez2024} follows the framework established by \citet{grossschedl2019}, which is based on the observational classes originally defined by \citet{lada1987} and also adopted by \citetalias{richardson24} in their classification, as given in Equation~\ref{spectral_index}. The only distinction from the \citetalias{richardson24} classification is that \citet{hernandez2024} further subdivides the Class III phase into two groups: those with a disk and those without a disk. The classification scheme used by \citet{hernandez2024} is as follows:

\begin{itemize}
    \item \textbf{0/I}: \(\alpha \geq 0.3\)
    \item \textbf{Flat}: \(-0.3 \leq \alpha < 0.3\)
    \item \textbf{II}: \(-1.6 \leq \alpha < -0.3\)
    \item \textbf{III thin disk}: \(-2.5 \leq \alpha < -1.6\)
    \item \textbf{III no disk / MS}: \(-2.5 \geq \alpha\)
    \label{spectral_index}
\end{itemize}

\begin{figure}
\centering
\includegraphics[scale=0.3]{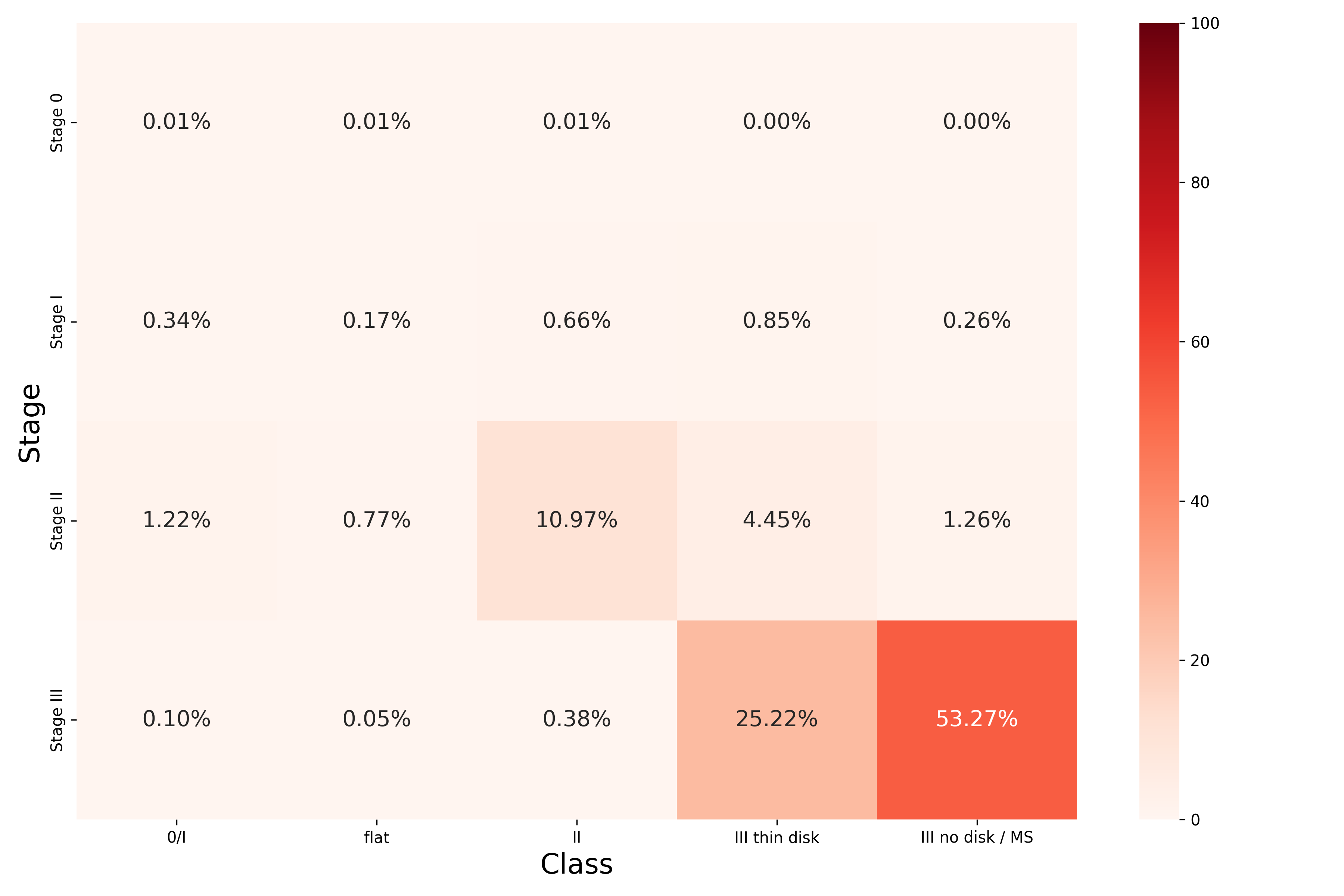}
\caption{The heatmap showing the distribution of stellar sources across different Stage and Class categories. Stages are derived from SED fitting, while Classes are adopted from \citet{hernandez2024}, providing a visual representation of the relationship between stellar evolutionary stages and classifications.}
\label{stage_class_matrix}
\end{figure}

In Fig.~\ref{stage_class_matrix}, we present the distribution of sources across different combinations of SED-derived \textit{Stage} classifications and the \textit{Class} assignments from \citet{hernandez2024}. It is often assumed that the \textit{Class} and \textit{Stage} classification systems are aligned, with \textit{Class 0/I} corresponding to \textit{Stage 0}, \textit{Flat} to \textit{Stage I}, \textit{Class II} to \textit{Stage II}, and \textit{Class III thin disk} and \textit{Class III no disk / MS} to \textit{Stage III}, representing equivalent evolutionary phases.

Fig.~\ref{stage_class_matrix} reveals a strong correlation between \textit{Class III no disk / MS} and \textit{Class III thin disk} with \textit{Stage III}, as well as between \textit{Flat} and \textit{Stage II}. This observed alignment suggests that the \textit{Class}-based and \textit{Stage}-based classifications exhibit a high degree of consistency, particularly in more evolved phases such as \textit{Stage III} and \textit{Class III no disk / MS}. 

The relationship between these two classification systems indicates that the \textit{Stage} classification effectively captures the physical evolution of YSOs, while the \textit{Class} classification reflects observable characteristics, such as disk structure and envelope material. This alignment underscores the utility of both classification frameworks for tracking YSO evolution, with the \textit{Stage} classification providing insights into physical processes and the \textit{Class} classification highlighting observable features.

\subsection{Analysis of Evolutionary Trends in the OSFC}
Circumstellar disks are critical structures in the star formation process, originating from the conservation of angular momentum during the collapse of molecular clouds. Advances in high-resolution instruments, such as ALMA \citep{ALMA09}, have enabled detailed studies of these disks, revealing features such as Keplerian rotation and ring-like substructures \citep{tobin16a, tychoniec18, tobin2020}. These features are believed to play a pivotal role in early dust evolution and planet formation.

Theoretical models of disk formation integrate advanced physical processes, such as non-ideal magnetohydrodynamics, turbulence, and radiation transfer \citep{zhao20, manara18, manara23, yen24}. Bridging these theoretical advancements with high-resolution observational data is essential for a comprehensive understanding of star and disk formation. While the debate about whether disks can form early in the star formation process may be resolved, the properties of actual protostellar disks remain critical for constraining these simulations.

Disk parameters, such as disk mass, cannot be well constrained without detections in the sub-mm; mid-infrared fluxes only provide a lower limit on the disk mass \citepalias{robitaille2017}. However, given that our catalog provides large number of derived stellar parameters, we analyzed our dataset to investigate disk properties and to constrain the timescales of disk formation and evolution.

\begin{figure}
\centering
\includegraphics[scale=0.30]{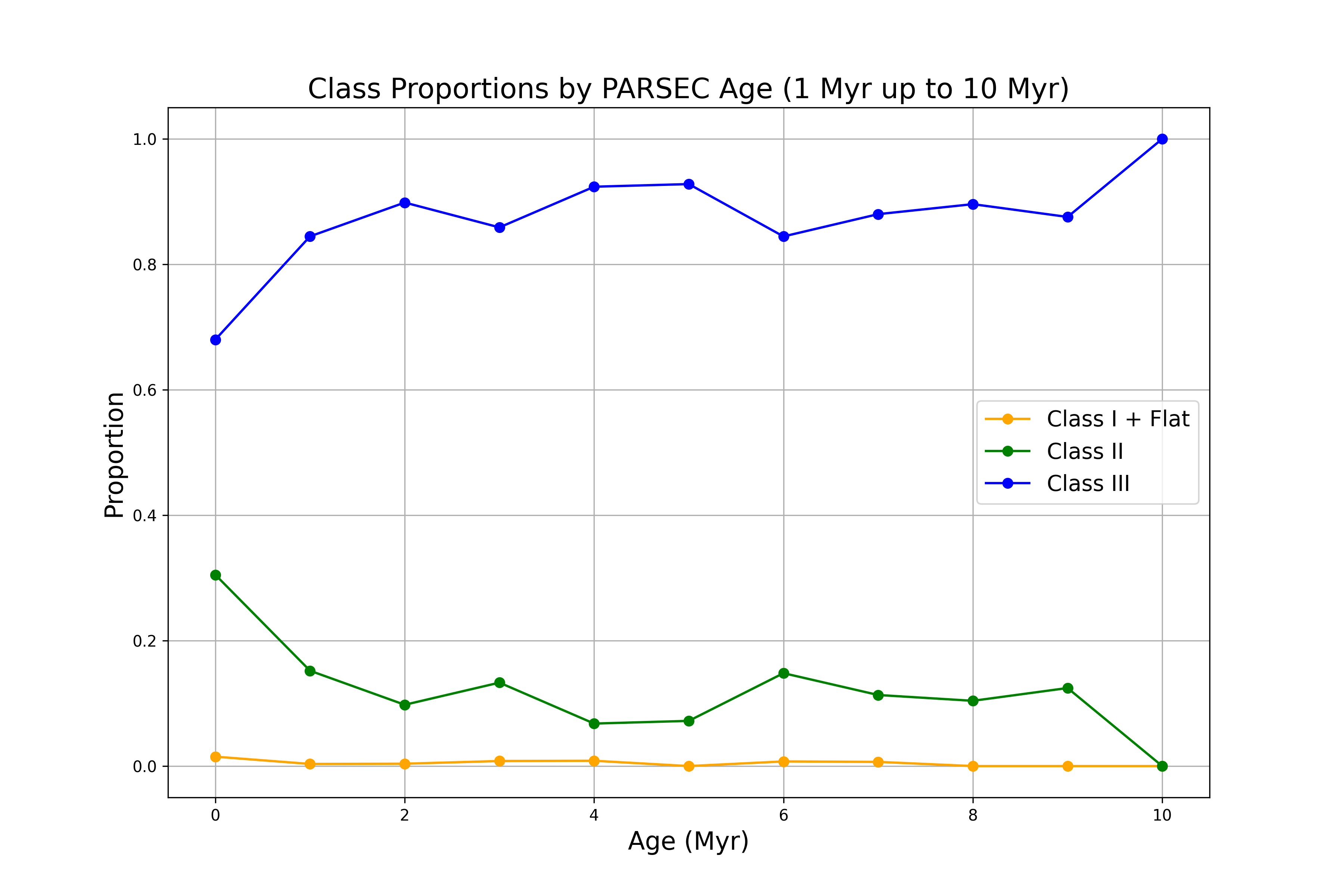}
\caption{Class proportions by age using uniform 1 Myr bins for sources in the PARSEC age range of 0–10 Myr. The orange line (Class I + Flat) shows consistently low values due to the scarcity of these evolutionary stages in the dataset. Uniform bins were used for consistency and ease of comparison across the age range. The fractions were calculated as the number of sources in each class divided by the total number of sources in the respective age bin.}
\label{class_age}
\end{figure}

To investigate the evolutionary trends of stellar populations in the OSFC, we analyzed the distribution of stellar classes (Class I, Flat, Class II, and Class III) as a function of age, 
and we restricted the analysis to stars with age estimates up to 10 Myr. In Fig.~\ref{class_age}, we present the class fractions as a function of age for the best-fit solutions of 15,396 sources. Stellar ages were grouped into 1 Myr bins to capture trends in class proportions across discrete intervals. We retained uniform bins to maintain consistency across the entire age range and simplify direct comparisons between different stages of evolution.

Class I and Flat stars represent young sources with significant circumstellar material, indicative of early evolutionary phases. Class II stars, characterized by protoplanetary disks, represent an intermediate stage in stellar evolution, while Class III stars exhibit minimal or no circumstellar material, marking the transition to main-sequence-like behavior. This analysis confirms the expected decline in the number of Class I + Flat and Class II sources with increasing age, consistent with stellar evolution models. The consistently low fraction of Class I + Flat sources reflects the advanced evolutionary state of many objects in this dataset.

\subsection{Mass-Luminosity Relation}
\label{mass-luminosity}
\begin{figure}
\centering
\includegraphics[scale=0.35]{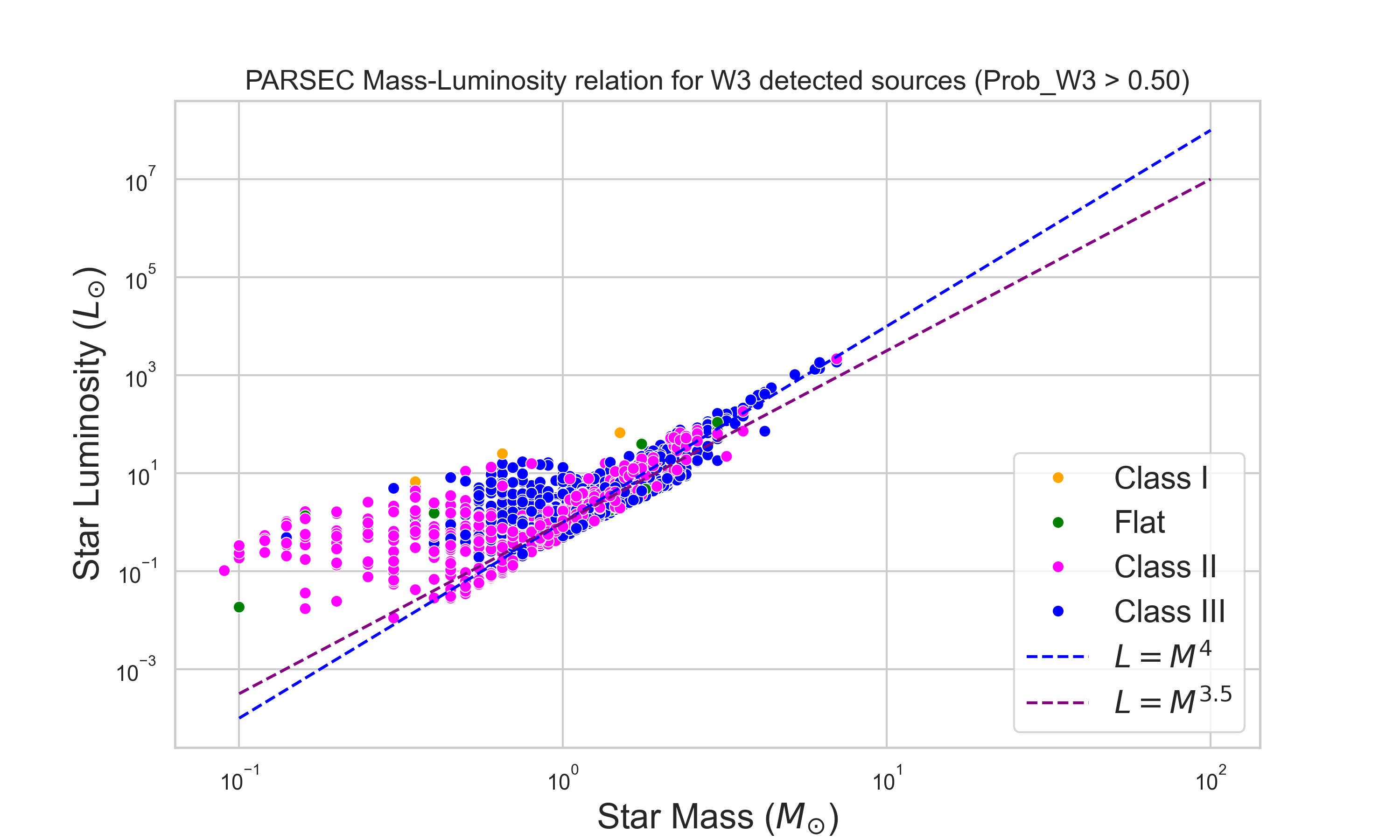}
\caption{Mass-Luminosity plot is presented. This shows mass calculated from PARSEC evolutionary tracks (with solar metallicity) vs luminosity derived from the SED fitting of sources. The colors of the data points represent different classes.}
\label{mass_lum}
\end{figure}

The mass-luminosity relation was also analyzed using the calculated masses from the PARSEC evolutionary tracks. Fig.~\ref{mass_lum} shows the masses for 5,062 high-confidence sources that fall within the range of the PARSEC evolutionary tracks (solar metallicity). A clear upward trend is seen, indicating that as the mass of the stars increases, their luminosity also increases, consistent with the mass-luminosity relation. The relationship appears to be approximately linear for stars with masses between 0.5 and 2 M$_\odot$, consistent with the relation \(\left(\frac{L}{L_\odot}\right) = \left(\frac{M}{M_\odot}\right)^4\). For higher masses, likely above 2 M$_\odot$, the slope begins to change, suggesting a transition to a region where the exponent becomes approximately 3.5, which corresponds to the main-sequence relation.

As the mass increases, the number of observed stars decreases, particularly for very high masses. This trend is consistent with the understanding that massive stars are less common, and evolve to the main sequence on a shorter timescale.

For low-mass stars, particularly those with masses less than approximately 0.5 M$_\odot$, the relationship between mass and luminosity becomes less straightforward. The study of stellar masses and PMS evolution has seen significant advancements, with several works evaluating the accuracy of different mass determination methods and model predictions \citep{HW04,bell12, bell13}. \cite{HW04} compared dynamical and evolutionary mass determination techniques using a sample of 115 stars, including both main-sequence and PMS stars. Their results showed generally good agreement for stars with masses above 1.2 M$_\odot$, but for lower-mass stars (down to 0.3 M$_\odot$), evolutionary models consistently underpredicted the masses by 10-30\%, with discrepancies as large as 50-100\% for stars below 1.0 M$_\odot$. This difference was attributed to inaccuracies in the treatment of convection within the evolutionary models. A similar trend can be observed in Fig.~\ref{mass_lum}, where stars with masses less than 1 M$_\odot$ show higher luminosities (calculated from SED fitting) and lower masses (derived from stellar evolutionary tracks), highlighting the mass-luminosity mismatch.

Regarding classification, Class I objects are expected to exhibit high luminosities, primarily originating from accretion rather than intrinsic stellar emission. While the Stefan-Boltzmann law provides a reasonable estimate of luminosity based on effective temperature (\(T_{\text{eff}}\)), the model assumption may not fully capture the complexities of such systems, particularly when circumstellar material is significant. For Class II objects, the observed luminosity is predominantly stellar in origin but remains higher than the ZAMS luminosity, consistent with their pre-main-sequence evolutionary stage. Given the likely contribution from circumstellar material (e.g., a disk), the derived effective temperature (\(T_{\text{eff}}\)) and radius primarily reflect the stellar component, resulting in an estimate of stellar luminosity rather than the total bolometric luminosity.

The plot effectively illustrates the mass-luminosity relation and provides valuable insights into different stellar classes. The upward trend in luminosity with increasing mass for sources with masses greater than 1 M$_\odot$ is a hallmark of stellar physics and aligns with the empirical relations observed in astrophysics \citep{kippenhahn94, tout96, torres2010}.

\subsection{FIR/Sub-mm Observations and the Spatial Distribution of YSOs in the OSFC}
\label{spatial-distribution}

Previous far-infrared (FIR) studies of the OSFC using Spitzer and other facilities have considerably advanced our understanding of star formation in this region. For instance, \cite{megeath12, megeath16} and \cite{kounkel18} have mapped the distribution of dusty young stellar objects and provided detailed classifications of their evolutionary stages. These works reveal that different regions within the OSFC exhibit distinct characteristics. Specifically, Orion A—particularly around the Orion Nebula Cluster (ONC)—is dominated by very young, embedded protostars (Class I and Flat-spectrum sources), whereas Orion B tends to host a larger fraction of more evolved objects. FIR observations are critical in this context because they detect the dusty envelopes and circumstellar disks that are otherwise challenging to observe, thereby reducing the observational biases that could lead to an underestimation of early-stage star formation.

\begin{figure*}
\centering
\includegraphics[scale=0.50]{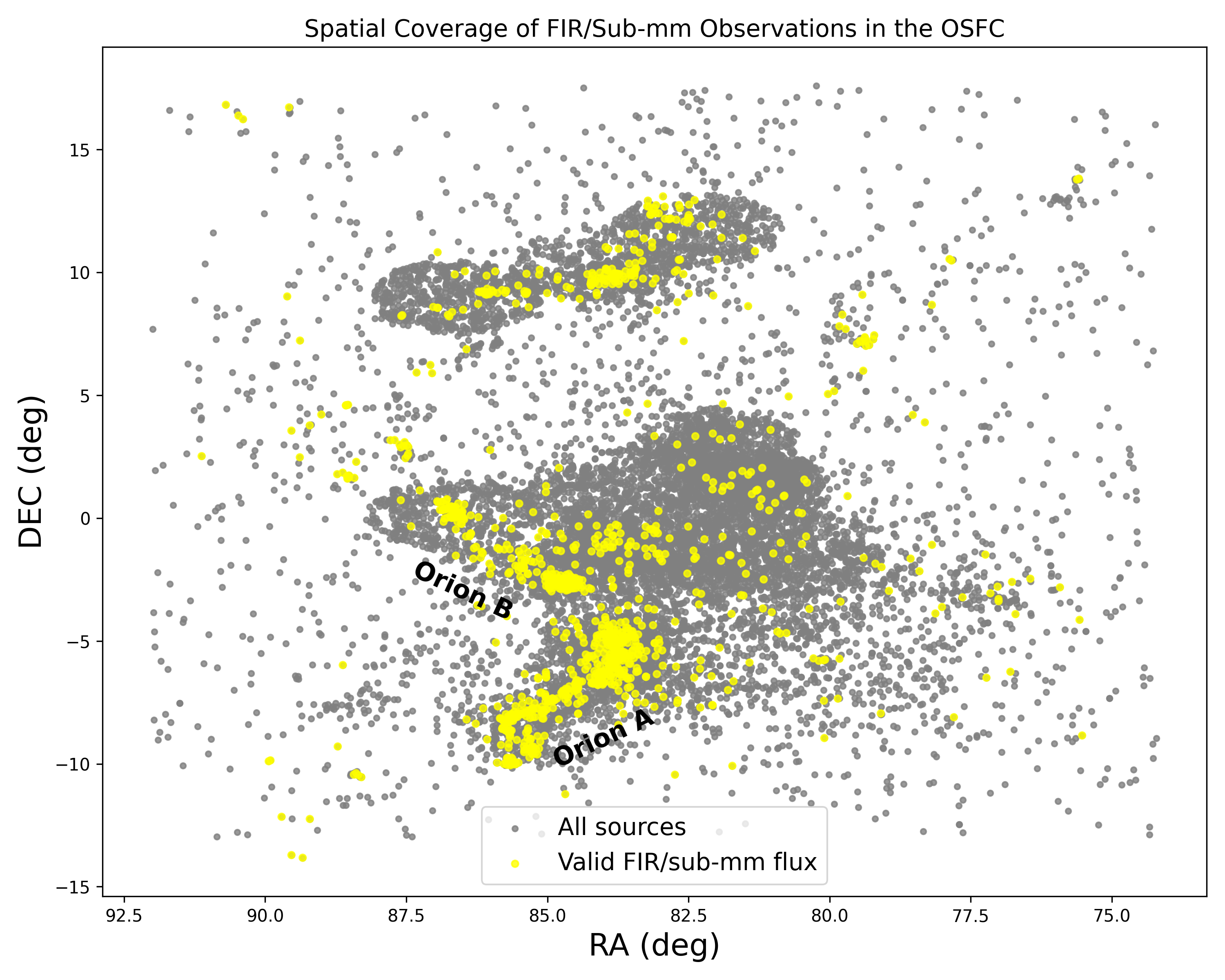}\
\includegraphics[scale=0.50]{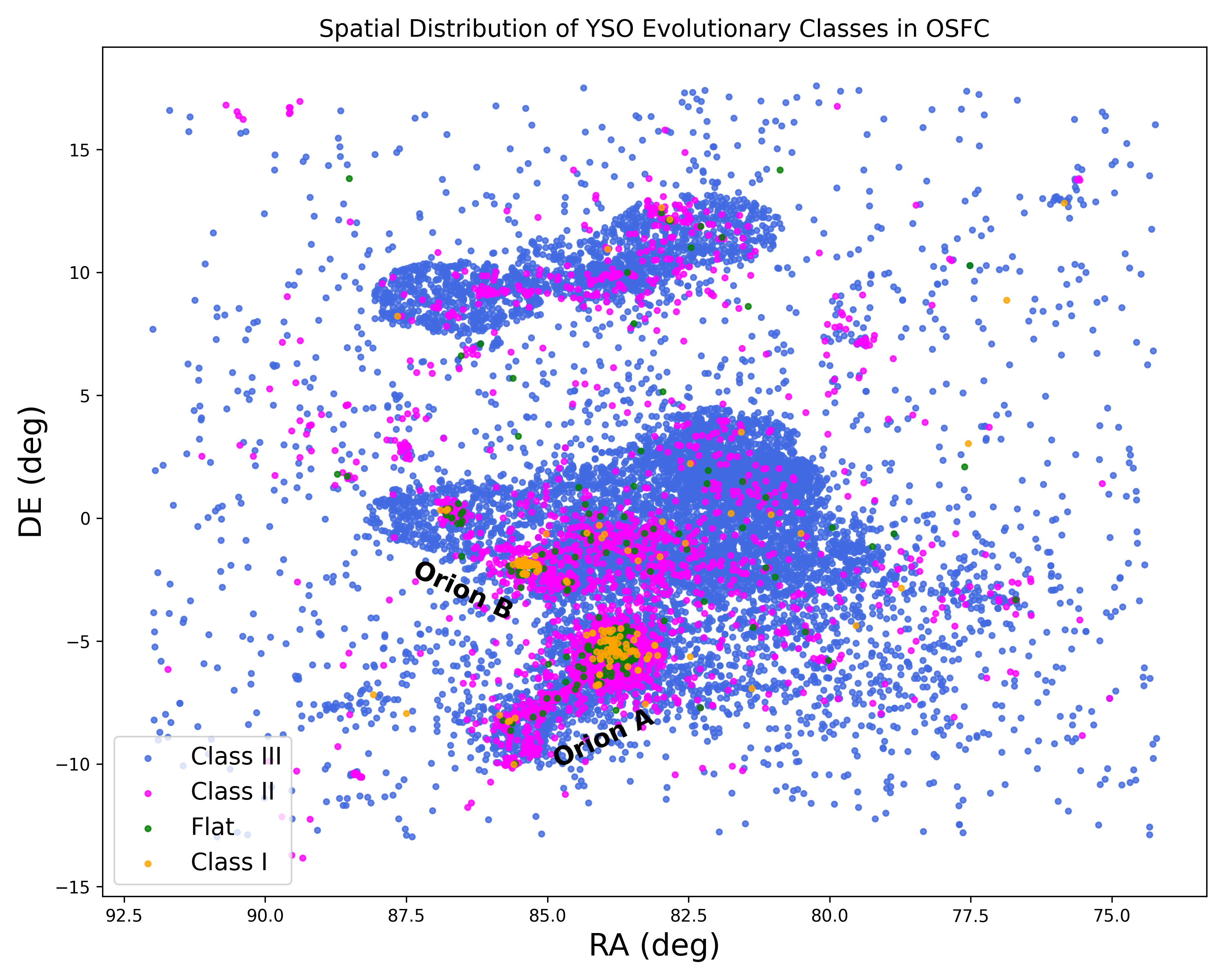}
\caption{The upper panel shows the observational coverage in the FIR/Sub-mm wavelengths. The lower panel illustrates the spatial distribution of young stellar objects (YSOs) categorized by their evolutionary classes.}
\label{flux_spatial_coverage}
\end{figure*}

In the upper panel of Figure~\ref{flux_spatial_coverage}, we present the spatial distribution of YSOs in the OSFC in RA/Dec coordinates. Gray points denote all detected sources, while yellow points highlight those with at least one valid flux measurement in the FIR/Sub-mm filters (including \textit{AllWISE}/W4, \textit{Spitzer}/MIPS, \textit{Herschel}/PACS, and APEX/LABOCA/SABOCA). Expressing our data in RA/Dec allows for direct comparison with published observational footprints, such as Figure~1 in the Spitzer survey of the Orion A \& B molecular clouds \cite{megeath12}.

The lower panel of Figure~\ref{flux_spatial_coverage} shows the spatial distribution of YSO classes derived from our SED fitting across the OSFC. Notably, the regions covered by FIR/Sub-mm observations largely coincide with areas where the youngest objects (Class I and Flat-spectrum sources) are concentrated. This finding is consistent with Spitzer observations and underscores the importance of long-wavelength data for detecting protostellar envelopes and disks. In the absence of such observations, early-stage YSOs may be underestimated, potentially introducing an observational bias. Alternatively, regions lacking Class I and Flat-spectrum sources might be genuinely more evolved, with little circumstellar material remaining. Distinguishing between these possibilities requires uniform, high-resolution FIR/Sub-mm coverage of the entire complex.

Given that our study incorporates the most comprehensive archival data currently available, our findings are in strong agreement with previous results, reinforcing the established picture of the OSFC's star-forming regions. While our analysis does not reveal fundamentally new structures, it provides an independent and updated confirmation of the spatial distribution of YSOs and their evolutionary stages, further validating the conclusions drawn from earlier surveys. Our results highlight the importance of continued observational efforts, particularly in the FIR/Sub-mm regime, to refine our understanding of the OSFC’s complex star formation history.

\section{Summary}
\label{conclusion}
In this study, we have performed a comprehensive SED fitting analysis of YSOs, utilizing a dataset that spans 15,396 sources. The stellar parameters derived from this extensive analysis account for the varying completeness and reliability of the available data. Notably, when considering the more reliable subset of sources that exhibit at least a W3 (12 $\mu$m) detection at longer wavelengths, with a probability (\textit{Prob\_W3}) greater than 50\%, we identify a total of 5,062 sources. 

The integration of multi-wavelength data is crucial for a more nuanced understanding of star formation. 
Looking forward, we are incorporating advanced machine learning and deep learning techniques to assist in the identification and classification of YSOs based on their photometric data and light curves. Future work will involve comparing the results obtained through traditional SED fitting methods with those derived from the newly implemented machine learning and deep learning approaches. This comparison will help us assess the strengths and limitations of each method, ultimately refining our understanding of YSOs and their evolutionary processes.

\begin{acknowledgements}
  This work has received funding from the European Union's Horizon 2020 research and innovation program under grant number 101004141. G.M. acknowledges support from the János Bolyai Research Scholarship of the Hungarian Academy of Sciences. This research was supported by the International Space Science Institute (ISSI) in Bern, through ISSI International Team Project 521 selected in 2021, Revisiting Star Formation in the Era of Big Data (\url{https://teams.issibern.ch/starformation/}). IG expresses her thanks to Theo Richardson for his assistance with the updated SED model.
\end{acknowledgements}

%
%

\bibliography{ilk_lib}
\newpage

\begin{appendix}

\section{Example SED fitting results}\label{SED_18}

\begin{sidewaystable*}
    \centering
    \caption{Examples of SED fitting results}
    \label{selected_rows}
    \resizebox{\textwidth}{!}{%
    \begin{tabular}{l l l l l l l l l l r r r r r r r l l l}
    \hline
        \textbf{InternalID} & \multicolumn{1}{c}{\textbf{RA}} & \multicolumn{1}{c}{\textbf{DEC}} & \textbf{ndpt} & \textbf{Flux Code} & \textbf{Prob W3} & \textbf{Prob W4} & \multicolumn{1}{c}{\textbf{Distance}} & \textbf{Model Set} & \textbf{nfits} & \textbf{$\chi^2_{\text{min}}$} & \textbf{R$_\star$} & \textbf{T$_\star$} & \textbf{L$_\star$} & \textbf{M$_\star$} & \textbf{Age} & \textbf{Inclination} & \textbf{Class} & \textbf{Stage} \\
                  & \multicolumn{1}{c}{\textbf{(deg)}} & \multicolumn{1}{c}{\textbf{(deg)}} &      &           &          &          & \multicolumn{1}{c}{\textbf{(pc)}}      &                  &                &                                    & \textbf{(R$_\odot$)} & \textbf{(K)}    & \textbf{(L$_\odot$)} & \textbf{(M$_\odot$)} & \textbf{(Myr)} & \textbf{($^\circ$)}    &          \\
    \hline
    ID7166 & 83.57055 & -6.54707 & 13 & M1 & 0.03 &   & 363.62 & spubsmi & 1 & 238.95 & 0.62 & 3523 & 0.05 & 0.55 & 55.7 & 28.04 & Class II & Stage II \\
    ID9189 & 85.06181 & -6.34854 & 10 & W4 & 1.00 & 1.00 & 368.06 & sp--s-i & 10 & 45.26 & 0.88 & 6044 & 0.93 & 1.05 & 36.1 & 27.56 & Class II & Stage II \\
    ID1738 & 79.22729 & -1.44311 & 6 & W3 & 0.99 &  & 311.13 & s---s-i & 21 & 39.5 & 1.61 & 5594 & 2.28 & 1.40 & 9.09 & 45.00 & Class III & Stage III \\
    ID26196 & 85.12050 & 9.24075 & 6 & W3 & 0.56 &  & 400.0\footnotemark[1] & s---s-i & 11 & 23.5 & 1.11 & 4506 & 0.46 & 0.95 & 8.50 & 45.00 & Class III & Stage III \\
    ID4996 & 83.18252 & -0.45422 & 9 & P3 & 0.98 & 1.00 & 326.71 & sp--hmi & 1 & 131 & 2.49 & 6661 & 11.01 & 1.70 & 7.89 & 5.01 & Class II & Stage II \\
    ID27106 & 86.91462 & 7.78719 & 6 & W3 & 0.67 &  & 400.0\footnotemark[1] & s---smi & 18 & 53.1 & 1.49 & 4844 & 1.10 & 1.25 & 5.35 & 45.00 & Class III & Stage III \\
    ID18481 & 82.69350 & 8.78547 & 6 & W4 & 0.94 & 0.98 & 397.38 & s-u-smi & 30 & 9.57 & 1.48 & 4475 & 0.79 & 1.00 & 3.86 & 9.34 & Class II & Stage I \\
    ID4488 & 85.79479 & -8.56539 & 9 & M1 & 0.99 & 1.00 & 400.0\footnotemark[1] & s-u-hmi & 1 & 85.4 & 1.48 & 3922 & 0.47 & 0.70 & 2.82 & 78.28 & Class II & Stage I \\
    ID378 & 85.36927 & 9.17952 & 7 & W4 & 1.00 & 0.97 & 397.63 & sp--s-i & 24 & 39.26 & 1.27 & 3196 & 0.15 & 0.30 & 2.12 & 64.78 & Class II & Stage II \\
    ID5163 & 84.82848 & -2.51479 & 14 & P3 & 1.00 & 1.00 & 409.63 & sp--smi & 10 & 156 & 1.82 & 3947 & 0.72 & 0.65 & 1.42 & 35.00 & Class II & Stage II \\
    ID20380 & 87.78089 & 3.16753 & 8 & P3 & 0.94 &  & 409.79 & s-ubsmi & 27 & 65.2 & 2.11 & 4571 & 1.75 & 1.00 & 1.30 & 21.18 & Class III & Stage I \\
    ID5098 & 84.51990 & -1.28853 & 7 & W4 & 0.99 & 1.00 & 346.54 & s-p-hmi & 1 & 23.9 & 1.52 & 3308 & 0.25 & 0.30 & 1.27 & 45.00 & Class II & Stage I \\
    ID1434 & 87.02274 & 8.59186 & 7 & W4 & 1.00 & 1.00 & 393.90 & s-p-smi & 1 & 54.9 & 2.05 & 3439 & 0.53 & 0.30 & 0.70 & 45.00 & Class II & Stage I \\
    ID20379 & 87.46348 & 2.88613 & 7 & W4 & 1.00 & 0.95 & 411.25 & s-pbhmi & 1 & 65 & 2.11 & 3610 & 0.68 & 0.35 & 0.63 & 85.66 & Class II & Stage I \\
    ID7239 & 85.3521 & -8.94225 & 15 & M1 & 1.00 & 1.00 & 465.0 & sp--hmi & 1 & 182.3 & 2.65 & 3966 & 1.56 & 0.5 & 0.43 & 55.33 & Class II & Stage II \\
    ID3525 & 88.42046 & 1.63725 & 10 & P3 & 1.00 & 1.00 & 334.92 & spubsmi & 3 & 114 & 3.29 & 3887 & 2.23 & 0.45 & 0.18 & 30.93 & Class II & Stage II \\
    ID4777 & 84.69388 & -7.09372 & 17 & L & 0.81 & 0.96 & 400.0\footnotemark[1] & spubhmi & 1 & 681 & 2.46 & 3314 & 0.66 & 0.25 & 0.17 & 52.74 & Class II & Stage I \\
    ID1447 & 80.72661 & 4.94878 & 7 & W4 & 1.00 & 1.00 & 350.46 & s-ubhmi & 2 & 60.4 & 2.41 & 3151 & 0.51 & 0.16 & 0.08 & 41.26 & Class II & Stage I \\
    ID9858 & 86.18356 & -1.61107 & 10 & P3 & 1.00 & 1.00 & 477.28 & sp--h-i & 3 & 99.7 & 3.46 & 2969 & 0.84 & 0.14 & 0.01 & 14.00 & Class II & Stage II \\
    ID9045 & 86.90238 & 0.33506 & 7 & L & 1.00 & 0.98 & 400.0\footnotemark[1] & spu-hmi & 1 & 21.5 & 10.24 & 4034 & 25.02 & 0.65 & 0.01 & 9.30 & Class I & Stage I \\
    \hline
    \end{tabular}%
    }
     \tablefoot{Example of 20 SED fitting results, sorted by age. The columns are as follows:(1) internal source identifier, (2) and (3) source coordinates in units of degrees, (4) number of data points used in the fit, (5) flux codes, which denotes the longest detected wavelength for each source (6) and (7) reliability probability of the \textit{WISE} W3 and W4 detections, (8) distance based on \textit{Gaia} DR3 parallax. Column (9) is the models used for fitting, (10) number of model fits attempted (11) (\(\chi^2\)) per data point for the best-fit model from the most likely model set, (12)--(17) list properties derived from the PARSEC models: radius, temperature, luminosity, mass, age, and inclination of the system. Column (18) corresponds to the class label, while column (19) lists the derived evolutionary stage. The distance 400.00\(^1\) (average OSFC distance) was adopted in cases where the parallax measurement was either unavailable or unreliable.}
\end{sidewaystable*}

\begin{figure*}
\vspace{20mm}
\centering
\includegraphics[scale=0.45]{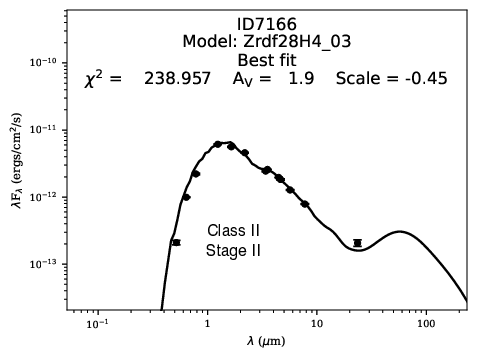}\
\includegraphics[scale=0.45]{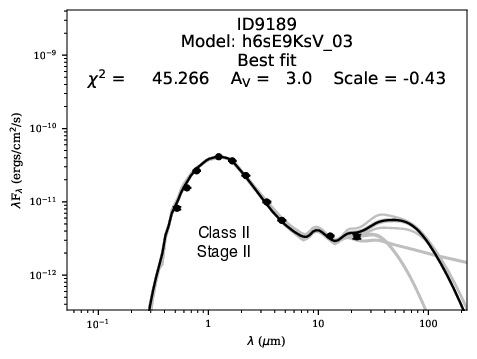}\\
\includegraphics[scale=0.45]{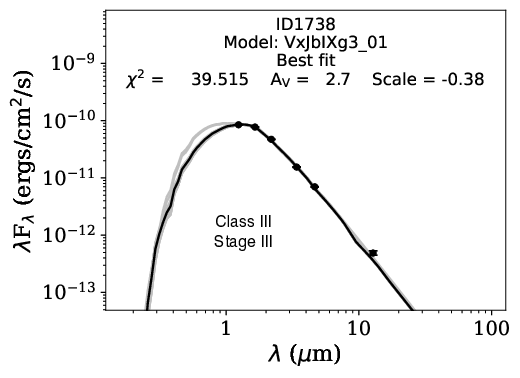}\
\includegraphics[scale=0.45]{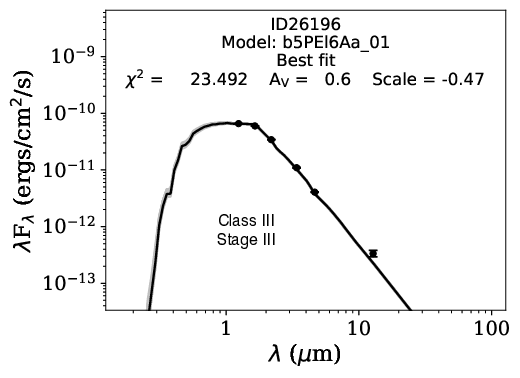}\\
\includegraphics[scale=0.45]{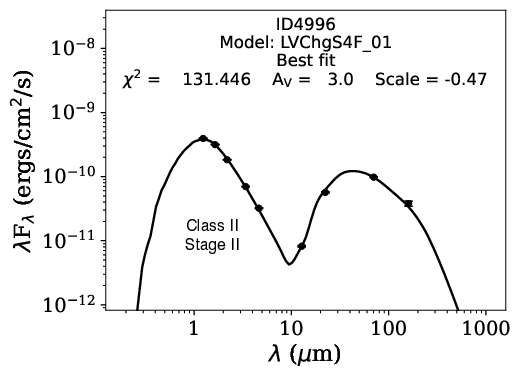}\
\includegraphics[scale=0.45]{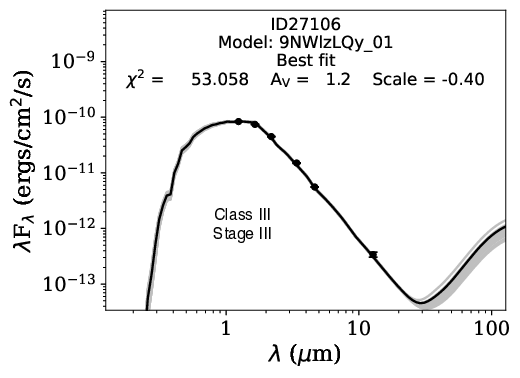}\\
\label{SEDs}
\end{figure*}

\begin{figure*}
\vspace{20mm}
\centering
\includegraphics[scale=0.45]{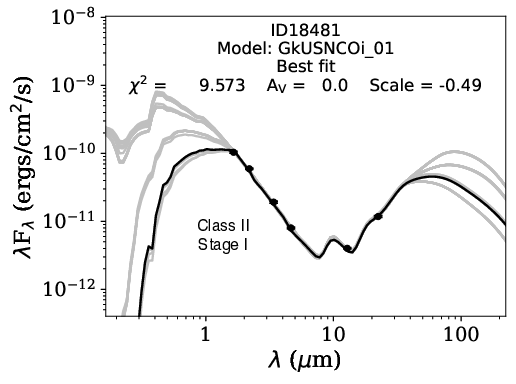}\
\includegraphics[scale=0.45]{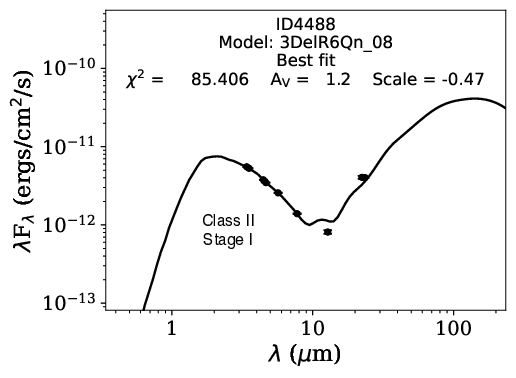}\\
\includegraphics[scale=0.45]{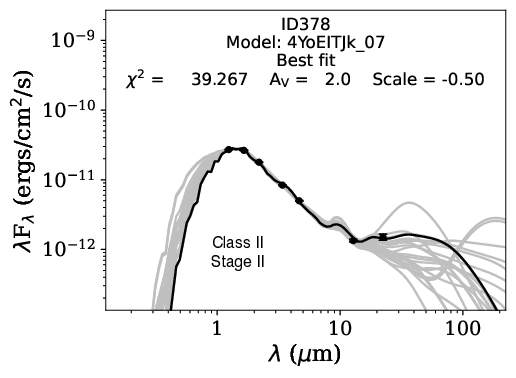}\
\includegraphics[scale=0.45]{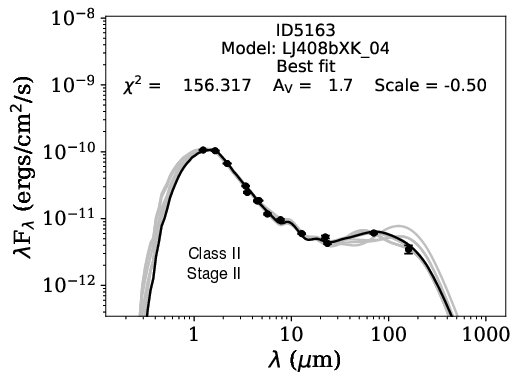}\\
\includegraphics[scale=0.45]{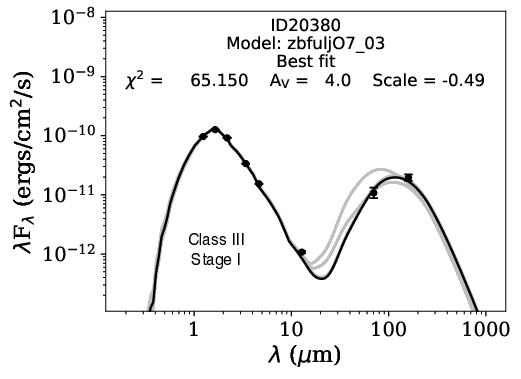}\
\includegraphics[scale=0.45]{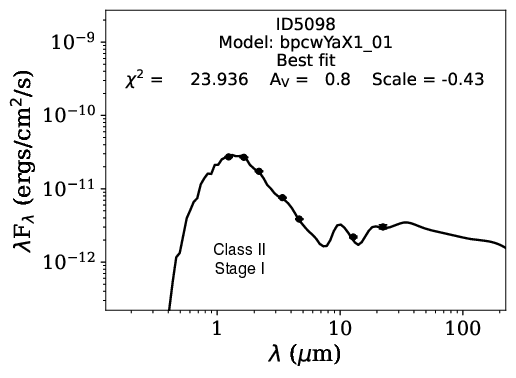}\\
\end{figure*}

\begin{figure*}
\centering
\vspace{20mm}
\includegraphics[scale=0.45]{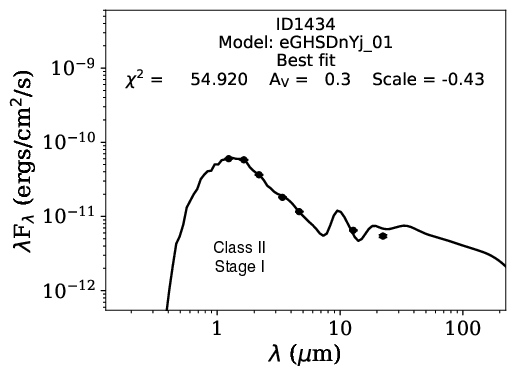}\
\includegraphics[scale=0.45]{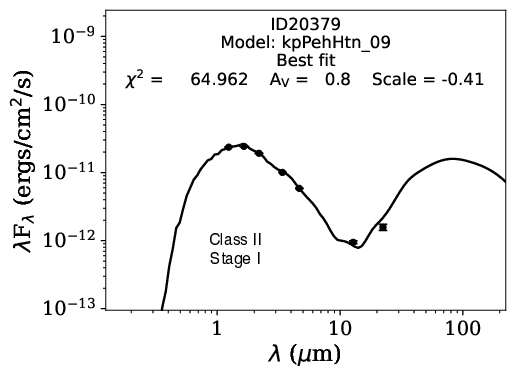}\\
\includegraphics[scale=0.45]{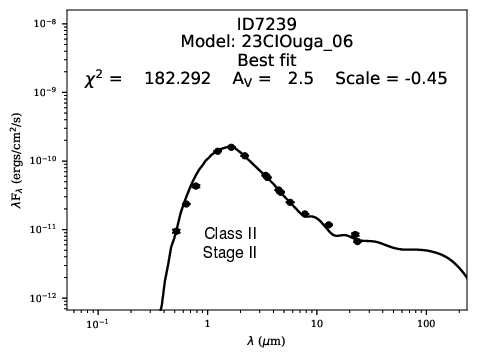}\
\includegraphics[scale=0.45]{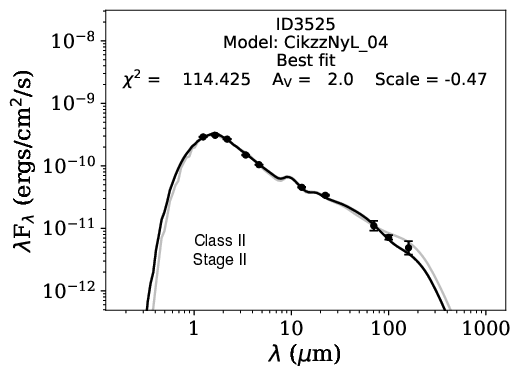}\\
\includegraphics[scale=0.45]{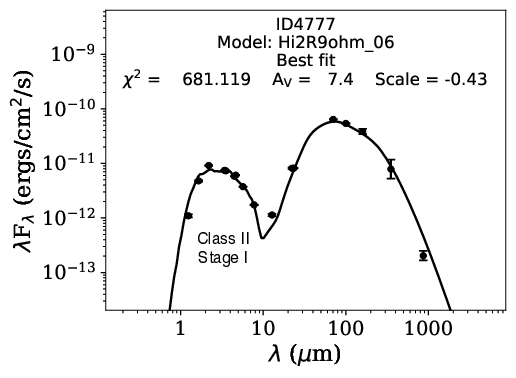}\
\includegraphics[scale=0.45]{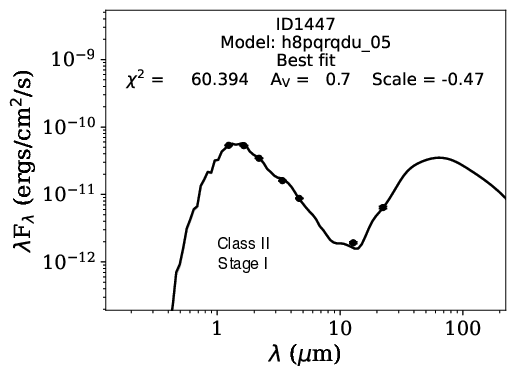}\\
\end{figure*}

\begin{figure*}
\vspace{20mm}
\centering
\includegraphics[scale=0.45]{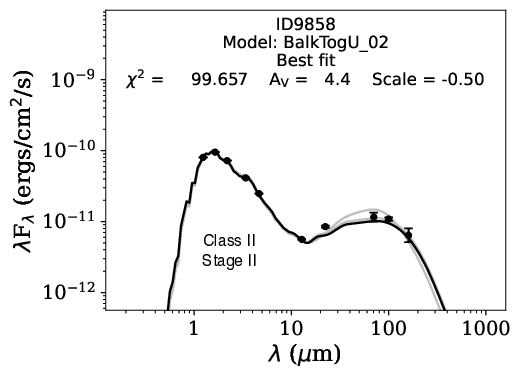}\
\includegraphics[scale=0.45]{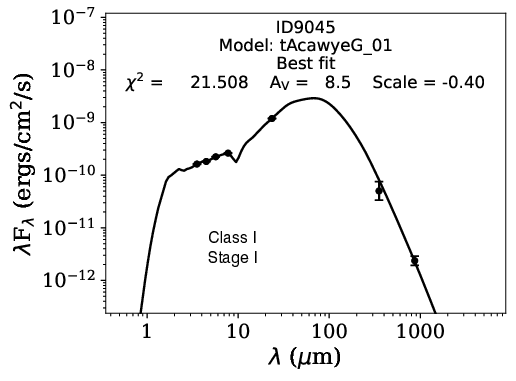}\\
\caption{The SED plots corresponding to the sources listed in Table \ref{selected_rows}. These examples illustrate sources at various evolutionary stages, effectively matched with the YSO models from \citetalias{robitaille2017}.}
\label{SEDs}
\end{figure*}

\end{appendix}
\end{document}